\documentclass[pra,aps,showpacs,notitlepage,superscriptaddress,tightenlines,twocolumn,raggedbottom]{revtex4-1}
\usepackage{graphicx,amsmath}
\usepackage{bbm}
\usepackage{txfonts}
\usepackage[utf8]{inputenc}
\usepackage{epstopdf}
\usepackage{siunitx}
\usepackage[colorlinks=true, citecolor=blue,linkcolor=blue,urlcolor=blue
]{hyperref}
\usepackage{color}
\usepackage{bm}
\usepackage{amsmath}
\usepackage{amsfonts}
\usepackage{amssymb}
\usepackage{graphicx}
\usepackage{epsfig}
\usepackage{epstopdf}
\usepackage{xcolor}
\usepackage{empheq}
\usepackage{cancel}
\usepackage{braket}
\DeclareMathOperator\arctanh{arctanh}

\newcommand{\ain}{\alpha_{\mathrm{i}}}
\newcommand{\cin}{\chi_{\mathrm{i}}}

\newcommand{\nae}{\bar{n}_{\mathrm{e,a}}}
\newcommand{\nai}{\bar{n}_{\mathrm{i,a}}}
\newcommand{\nce}{\bar{n}_{\mathrm{e,c}}}
\newcommand{\nci}{\bar{n}_{\mathrm{i,c}}}
\newcommand{\nm}{{\bar{n}_{\mathrm{m}}}}

\newcommand{\Ad}{A_{\mathrm{d}}}
\newcommand{\Ax}{A_{\mathrm{x}}}
\newcommand{\AdI}{A_{\mathrm{d,I}}}
\newcommand{\AxI}{A_{\mathrm{x,I}}}
\newcommand{\Am}{A_{\mathrm{m}}}
\newcommand{\Cd}{C_{\mathrm{d}}}
\newcommand{\Cx}{C_{\mathrm{x}}}
\newcommand{\CdI}{C_{\mathrm{d,I}}}
\newcommand{\CxI}{C_{\mathrm{x,I}}}
\newcommand{\Cm}{C_{\mathrm{m}}}

\begin{document}
\title{Clauser-Horne-Shimony-Holt Bell inequality test in an optomechanical device}

\author{Juuso~Manninen} 
\affiliation{Department of Applied Physics, Low Temperature Laboratory, Aalto University, P.O. Box 15100, FI-00076 AALTO, Finland}

\author{Muhammad~Asjad} 
\affiliation{Department of Physics and
  Nanoscience Center, University of Jyv{\"a}skyl{\"a}, P.O. Box 35 (YFL), FI-40014
  University of Jyv{\"a}skyl{\"a}, Finland}

%\affiliation{Department of Physics and
 % Nanoscience Center, University of Jyv{\"a}skyl{\"a}, P.O. Box 35 (YFL), FI-40014
  %University of Jyv{\"a}skyl{\"a}, Finland}
% \author{Elli Selenius} 
% \affiliation{Department of Physics and Nanoscience Center, University of Jyv{\"a}skyl{\"a}, 
% P.O. Box 35 (YFL), FI-40014 University of Jyv{\"a}skyl{\"a}, Finland}

\author{Risto~Ojaj{\"a}rvi} 
\affiliation{Department of Physics and Nanoscience Center, University of Jyv{\"a}skyl{\"a}, 
P.O. Box 35 (YFL), FI-40014 University of Jyv{\"a}skyl{\"a}, Finland}
  
\author{Petri~Kuusela} 
\affiliation{Department of Physics and Nanoscience Center, University of Jyv{\"a}skyl{\"a}, 
P.O. Box 35 (YFL), FI-40014  University of Jyv{\"a}skyl{\"a}, Finland}

\author{Francesco~Massel}
\email[]{francesco.p.massel@jyu.fi} \affiliation{Department of Physics and Nanoscience Center, University of Jyv{\"a}skyl{\"a}, 
P.O. Box 35 (YFL), FI-40014 University of Jyv{\"a}skyl{\"a}, Finland}

\begin{abstract}
  We propose here a scheme, based on the measurement of quadrature phase
  coherence, aimed at testing the Clauser-Horne-Shimony-Holt Bell inequality in
  an optomechanical setting. Our setup is constituted by two optical cavities
  dispersively coupled to a common mechanical resonator. We show that it is
  possible to generate EPR-like correlations between the quadratures of the
  output fields of the two cavities, and, depending on the system parameters, to
  observe the violation of the  Clauser-Horne-Shimony-Holt inequality.
\end{abstract}
\maketitle

\section{Introduction}
In his seminal work, motivated by the work by Einstein, Podolsky and Rosen,
\cite{Einstein:1935hx}, Bell showed that theories relying on local (possibly
hidden) variables, which are bound to satisfy certain inequalities, cannot
describe all quantum mechanical predictions \cite{Bell:1964wu}. From the point
of view of quantum theory, a violation of these Bell inequalities (BIs)
necessarily implies entanglement between spatially separated subsystems
\cite{Brunner:2014kr}. Beyond their intrinsic conceptual relevance, BI tests
have potentially important technological repercussions, allowing to certify the
security of quantum cryptographic schemes \cite{Acin:2007db}, making it relevant
to explore the possibility of performing such test in different setups and for
different physical systems.

Since the work of Bell, multiple experimental realizations of BI tests have been
conducted \cite{Freedman:1972ka,Fry:1976hv,Aspect:1982ja,Aspect:1982br,Weihs:1998cc,
  Rowe:2001ic,Matsukevich:2008bm,Ansmann:2009eq,Giustina:2013jsa,Christensen:2013dn,Hensen:2015dw,
  Giustina:2015fc, Shalm:2015cw, Thearle:2018ib}, the first one being performed
by Freedman and Clauser \cite{Freedman:1972ka}. However, the confirmation that,
without any additional assumptions --i.e.,  closing all loopholes--, predictions
offered by locally realistic theories cannot reproduce the experimental results
has been obtained only in the last few years \cite{Hensen:2015dw, Giustina:2015fc,
  Shalm:2015cw}. Even more recently, based on an early theoretical proposal
\cite{Huntington:2001kr} and resorting to  an experimental setup
similar to the employed in the Bell test performed by Ou and Mandel
\cite{Ou:1988gk}, a BI test relying on continuous variable measurement
has been performed \cite{Thearle:2018ib}.

Owing to the recent progresses in the concomitant manipulation of mechanical and
optical degrees of freedom at the quantum level \cite{Hammerer:2014vu,
  Aspelmeyer:2014ce}, cavity optomechanical systems represent one of the
cornerstones for future quantum information and communication technologies. On a
more fundamental level, these systems represent one of the most promising
platforms for experimental verification of physical theories, with applications
ranging from gravitational wave detection \cite{Abbott:2016ki} to the potential
observation of quantum gravitational effects \cite{Pikovski:2012hp} and
entanglement between nearly-macroscopic mechanical objects
\cite{Wang:2013hk, Woolley:2014he,  Massel:2017jx,Riedinger:2017wa,OckeloenKorppi:2018ks} .

In this spirit, in this article, we investigate the test of the
Clauser-Horne-Shimony-Holt (CHSH) \cite{Clauser:1969ff} BI in an optomechanical
system. Our main focus is a two-cavities optomechanical setup, either in the
microwave or in the visible-light regime, allowing for unrivaled flexibility in
the choice of detectors and transmission lines for loophole-free tests. In
addition, the nature of the optomechanical interaction characterizing our
proposal opens up the possibility for BI tests in mixed microwave/optical
settings \cite{Barzanjeh:2012ez} . The two cavities/one mechanics setup, which
we consider here for the BI test, was discussed in the past in connection with
entanglement properties of optomechanical systems \cite{Paternostro:2007hm,
  Wang:2013hk,Barzanjeh:2012ez} and was experimentally realized in the context
of multimode quantum signal amplification of microwaves
\cite{OckeloenKorppi:2016ke}. While other ideas for testing BIs in an
optomechanical setting have recently been
proposed\cite{Vivoli:2016fp,Hofer:2016id}, they are based on a rather different
setup than the one discussed here, for which, due to the sequential nature of the
pusling scheme, closing all loopholes, in particular the locality loophole,
requires to address extra technical difficulties as discussed in the
supplemental material of Ref. \cite{Vivoli:2016fp} which are not present in the
setup discussed here.  On more general grounds, it is worth mentioning that
closing the locality loophole in a microwave setting represents a formidable
challenge due to the necessity of the noiseless distribution of microwave
signals. In this sense an all-optical realization of our proposal would thus
seems favorable. In the following, however, in order to underline the relation
to the present state-of-the-art experimental capabilities, we mainly focus on the
experimental parameters of the microwave setup discussed in Ref.
\cite{OckeloenKorppi:2016ke}.

While the previous BI tests mentioned above rely either on the polarization
degree of freedom of optical photons
\cite{Freedman:1972ka,Fry:1976hv,Aspect:1982ja,Aspect:1982br,Weihs:1998cc,
  Giustina:2013jsa,Christensen:2013dn, Giustina:2015fc, Shalm:2015cw}, or on
different realizations of two-level systems in a condensed-matter context
\cite{Rowe:2001ic,Matsukevich:2008bm,Ansmann:2009eq,Hensen:2015dw}, our proposal
follows the ideas suggested by Tan \textit{et al.} \cite{Tan:1990ea,Tan:1991cg},
and considers the possibility of a CHSH BI violation through the detection of
the quadrature phases, in our case, in an optomechanical setting.

The paper is organized as follows. In Sec.~\ref{intro} we introduce the model
and discuss the conditions for the violation of the CHSH BI. In Sec.~\ref{RD} we
describe the numerical results for the violation of the BI and we show its
sensitivity to variations of other system parameters.  Lastly, we discuss the
effect of various noise sources on the violation of the inequality in our setup.

\section{Model and equation of motion }\label{intro}
The setup considered here is constituted by two electromagnetic resonant
cavities (A and C, respectively) --either in the optical or microwave regime--
dispersively coupled to a mechanical resonator. Following the standard
description of optomechanical systems \cite{Law:1995it, Genes:2009cb, Milburn:2012cu, Aspelmeyer:2014ce, Bowen:2015gt},
\begin{figure}[htb]
  \centering
  \includegraphics[width=\linewidth]{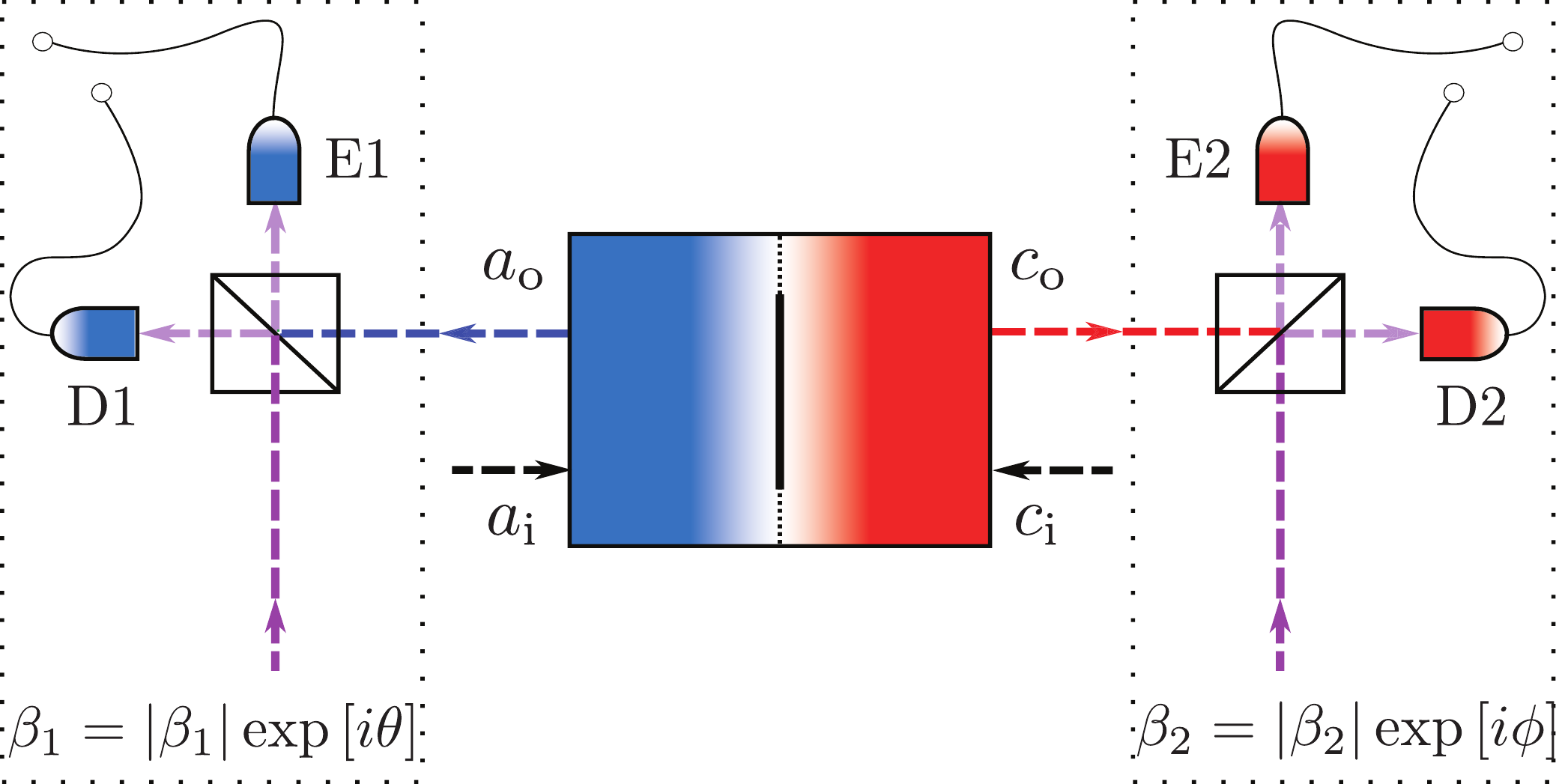}
  \caption{Schematic of the detection scheme. Outputs of the cavities are
    directed to different beam splitters, where they are mixed with local
    oscillators (LOs) fields. The mixed signals are sent to photodetectors D1,
    E1, D2, and E2, characterized by fields $d_1$, $e_1$, $d_2$, $e_2$,
    respectively.  Unlike the case of (balanced) homodyne detection schemes,
    where the signals emerging from the two branches of each beam splitter --in
    our case directed towards detectors D1/E1 and D2/E2-- are combined, we keep
    track of all four signals and their intensity correlations described by
    Eqs.(\ref{eq:21}-\ref{eq:24}).}
    \label{fig:1}
\end{figure}
 the Hamiltonian for the system can be written 
as
\begin{align}
  H=&\omega_\mathrm{a} a^\dagger a + \omega_\mathrm{c} c^\dagger c + \omega_\mathrm{m} b^\dagger b \nonumber \\
      &+ \left(g_\mathrm{a} a^\dagger a + g_\mathrm{c} c^\dagger c \right) \left(b^\dagger + b \right),
  \label{eq:1}
\end{align}
where $a$, $c$ and $b$ represent the lowering operators associated with cavity A
and C and the mechanical modes, respectively; $\omega_\mathrm{a}$,
$\omega_\mathrm{c}$, $\omega_\mathrm{m}$ are their resonant frequencies and
$g_\mathrm{a}$ and $g_\mathrm{c}$ are the single-photon radiation pressure
couplings for modes $a$ and $c$ with the mechanical mode.  
%We will deal with only a single mechanical mode even though other modes could be excited \cite{Pinard1999}, and the cavity spectral range will be chosen such that the scattering of photons into other optical modes is negligible \cite{Law1995}.

Along the lines of the experiment discussed in \cite{OckeloenKorppi:2016ke}, we
assume that each cavity is driven by a strong coherent tone
$\alpha_{\mathrm{in, a}}$ and $\alpha_{\mathrm{in, c}}$ (for cavity A and C, respectively). We consider that driving of each cavity is detuned
from the cavity resonance: we assume cavity A to be driven with a frequency
$\omega_{\mathrm{d,a}}=\omega_\mathrm{a} + \omega_\mathrm{m}$ (blue mechanical
sideband) and cavity C with a frequency
$\omega_{\mathrm{d,c}}=\omega_\mathrm{c} - \omega_\mathrm{m}$ (red mechanical
sideband). 
In our analysis, we employ the usual description of the system in
terms of quantum Langevin equations \cite{Walls:2008em} for the fluctuations
around the cavity fields induced by the drives. In this scenario, we consider
the linearized dynamics of the fluctuations around the pump tones and replace
$a \to a +\alpha_\mathrm{A}$ and $c \to c + \alpha_\mathrm{c}$ (see Appendix \ref{DM}).
%for the derivation of the equations
%of motion.
   %(%see %supplemental material (SM) \cite{supp} 
%(see also the SM).
Moving to a frame rotating at $\omega_{\mathrm{d,a}}$ and
$\omega_{\mathrm{d,c}}$ for modes $a$ and $c$ respectively and, defining
$\Delta_\mathrm{x}=\omega_{\mathrm{d,x}}-\omega_\mathrm{x}$ ($x=a,c$), we obtain
the following equations of motion for the fluctuations
\begin{subequations}
  \begin{align}
    \label{eq:2}
    \dot{a} =& \left(-i \Delta_\mathrm{a} -\dfrac{\kappa_\mathrm{a}}{2}\right)  a - i G_+  \left(b^\dagger + b \right)
                    +\sqrt{\kappa_\mathrm{e,a}}\,a_{\mathrm{i}}+ \sqrt{\kappa_\mathrm{i,a}}\,a_{\mathrm{I}},\\
    \label{eq:3}
    \dot{c} =& \left(-i \Delta_\mathrm{c}-\dfrac{\kappa_\mathrm{c}}{2}\right)  c -i G_-  \left(b^\dagger + b \right)
                    + \sqrt{\kappa_\mathrm{e,c}}\,c_{\mathrm{i}}+ \sqrt{\kappa_\mathrm{i,c}}\,c_{\mathrm{I}}, \\
    \label{eq:4}  
    \dot{b} =& \left(-\omega_\mathrm{m} -\dfrac{\gamma}{2}\right) b
                     - i G_+ \left(a^\dagger+ a\right) - i G_- \left(c^\dagger + c\right) + \sqrt{\gamma}\, b_{\mathrm{i}},
  \end{align}
\end{subequations}
where $G_+ = g_a \alpha_\mathrm{A} $ and $G_- = g_c \alpha_\mathrm{C}$ are the
linearized optomechanical couplings, and $\kappa_\mathrm{a}$,
$\kappa_\mathrm{c}$ and $\gamma$ are the linewidths of the cavities A, C and the
mechanical resonator. Moreover, we have defined $a_{\mathrm{i}}$,
$a_{\mathrm{I}}$, $c_{\mathrm{i}}$, $c_{\mathrm{I}}$, $b_{\mathrm{i}}$ to be the
input operators associated to the external input and internal fields
respectively ($i$ and $I$) for cavities A and C and the mechanics, respectively.

It is possible to obtain the expression of the cavity fields in frequency space
by Fourier transforming Eqs.~(\ref{eq:2}-\ref{eq:4}). The  transformation
leads to the following set of linear algebraic equations
\begin{subequations}
  \begin{align}
    \label{eq:5}
    -i \omega\, a =& \left(-i \Delta_\mathrm{a} -\dfrac{\kappa_\mathrm{a}}{2}\right)  a - i G_+  \left(b^\dagger + b \right)
                           +\sqrt{\kappa_\mathrm{e,a}}\,a_{\mathrm{i}}+ \sqrt{\kappa_\mathrm{i,a}}\,a_{\mathrm{I}},\\
    \label{eq:6}
    -i \omega\, c =& \left(-i \Delta_\mathrm{c}-\dfrac{\kappa_\mathrm{c}}{2}\right)  c -i G_-  \left(b^\dagger + b \right)
                           + \sqrt{\kappa_\mathrm{e,c}}\,c_{\mathrm{i}}+ \sqrt{\kappa_\mathrm{i,c}}\,c_{\mathrm{I}}, \\
    \label{eq:7}  
    -i \omega\, b =& -\dfrac{\gamma}{2} b - i G_+ \left(a^\dagger+ a\right) - i G_- \left(c^\dagger + c\right) + \sqrt{\gamma}\, b_{\mathrm{i}},
  \end{align}
\end{subequations}
which can be solved through standard techniques. Furthermore, according to the
input-output theory \cite{Walls:2008em}, the operators for the output fields of
cavity A are related to the cavity operators and to the input noise operators by
the relation $a_\mathrm{o}=\sqrt{\kappa_\mathrm{e,a}}\, a - a_\mathrm{i}$ where
$\kappa_\mathrm{e,a}$ is the external coupling rate for cavity A -- and
analogously for cavity C.

These relations, combined with the solution of Eqs.~(\ref{eq:5}\,- \ref{eq:7}),
allow us to map the the input cavity modes to the output fields $a_\mathrm{o}$,
$c_\mathrm{o}$ in the frequency domain as
\begin{subequations}
  \begin{align}
 \label{eq:8}
    a_{\mathrm{o}}=& A_\mathrm{d} a_{\mathrm{i}} +A_\mathrm{x} c^{\dagger}_{\mathrm{i}} +\mathcal{N}_\mathrm{a},\\
\label{eq:9}
    c_{\mathrm{o}}=& C_\mathrm{d} c_{\mathrm{i}} +C_\mathrm{x} a^\dagger_{\mathrm{i}} +\mathcal{N}_\mathrm{c}.     
  \end{align}
\end{subequations}
where the operators $\mathcal{N}_\mathrm{a}$ ($\mathcal{N}_\mathrm{c}$) account
for the noise associated with the mechanical resonator and the internal losses
of the cavity. In addition to these noise sources, we consider that the external
ports of the device represent potential further noise sources (see Appendix
\ref{IOeq}). % SM \cite{supp}).
% (see SM for further details).
While the direct solution of Eqs.~(\ref{eq:5}-\ref{eq:7}) outlined above is
sufficient to determine the value of the coefficients in Eqs.~\eqref{eq:8}, a
deeper physical intuition into the mechanism leading to the quantum correlations
among the modes --required for the violation of the BI-- can be obtained by
resorting to the rotating-wave approximation (RWA): the full derivation of the
expressions for the coefficients given in Eq.~\eqref{eq:8} within the RWA is
given in Appendix \ref{IOeq}, where we also compare RWA results with the full
solution of Eqs.~(\ref{eq:5}-\ref{eq:7}), which shows that, as it is usually the
case RWA and full results coincide in the so-called sideband resolved
regime($\omega_\mathrm{m}/\kappa \gg 1$).
% FPM add figure of the RWA comparison
% to the supplementary.
We outline here the key points of such derivation. In order to do this,
we write the EOMs in a frame rotating at the resonant frequency of
each mode 
\begin{subequations}
  \begin{align}
    \label{eq:10}
    \dot{a} =& -\dfrac{\kappa_\mathrm{a}}{2}    a
               - i G_+  \left(b^\dagger + b \exp\left[-2 i \omega_\mathrm{m} t\right]\right)
               + \sqrt{\kappa_\mathrm{e,a}}\,a_{\mathrm{i}}+ \sqrt{\kappa_\mathrm{i,a}}\,a_{\mathrm{I}},\\  
    \label{eq:11}
    \dot{c} =&-\dfrac{\kappa_\mathrm{c}}{2}     c
               - i G_-  \left(b^\dagger \exp\left[2 i \omega_\mathrm{m} t\right]+ b\right)
               +\sqrt{\kappa_\mathrm{e,c}}c_{\mathrm{i}}+\sqrt{\kappa_\mathrm{\rm  i, c}} c_{\mathrm{I}}, \\    
      \dot{b} =& -\dfrac{\gamma}{2}  b
               - i G_+  \left( a^\dagger+ a  \exp\left[-2 i \omega_\mathrm{m}t\right]  \right) \nonumber \\
               &\,\quad \quad - i G_-  \left(c+ c^\dagger  \exp\left[2 i \omega_\mathrm{m}t\right]\right) 
               + \sqrt{\gamma}\,b_{\mathrm{i}},   \label{eq:12}
  \end{align}
\end{subequations}
the RWA approximation consists in neglecting the (fast-rotating) time-dependent
terms in Eqs.~(\ref{eq:10}-\ref{eq:12}), leading to the following simplified
EOMs
\begin{subequations}
  \begin{align}
    \label{eq:13}
    \dot{a} =& -\dfrac{\kappa_\mathrm{a}}{2}    a - i G_+ b^\dagger  + \sqrt{\kappa_\mathrm{e,a}}\,a_{\mathrm{i}}+ \sqrt{\kappa_\mathrm{i,a}}\,a_{\mathrm{I}},\\  
    \label{eq:14}
    \dot{c} =&-\dfrac{\kappa_\mathrm{c}}{2}     c - i G_-   b  +\sqrt{\kappa_\mathrm{e,c}}c_{\mathrm{i}}+\sqrt{\kappa_\mathrm{\rm  i, c}} c_{\mathrm{I}}, \\    
    \label{eq:15}
    \dot{b} =& -\dfrac{\gamma}{2}  b - i G_+  a^\dagger   - i G_-  c+ \sqrt{\gamma}\,b_{\mathrm{i}}.
  \end{align}
\end{subequations}
We rewrite Eqs.~(\ref{eq:13}-\ref{eq:15}) in terms of two Bogolyubov operators
\begin{subequations}
  \begin{align}
    \label{eq:16}
    \eta_\mathrm{a}= \cosh \xi \, c + \sinh \xi \,a^\dagger ,\\
    \label{eq:17}
    \eta_\mathrm{c}= \cosh \xi \, a + \sinh \xi \,c^\dagger, 
  \end{align}
\end{subequations}
where $\cosh \xi=G_-/\mathcal{G}$, $\sinh \xi=G_+/\mathcal{G}$ with $\mathcal{G}=\sqrt{G^2_- - G^2_+}$ and rewrite Eq.~(\ref{eq:13}-\ref{eq:15}) in terms of the Bogolyubov modes $\eta_\mathrm{a}$ and $ \eta_\mathrm{c}$ as
\begin{subequations}
  \begin{align}
    \label{eq:18}
    \dot{\eta}_\mathrm{a} =& -\dfrac{\kappa}{2} \eta_\mathrm{a} - i \mathcal{G} b + \sqrt{\kappa_{\rm e}}  \eta_\mathrm{a,i}+ \sqrt{\kappa_{\rm i}}  \eta_\mathrm{a,I},\\
    \label{eq:19}
    \dot{\eta}_\mathrm{c} =& -\dfrac{\kappa}{2} \eta_\mathrm{c} + \sqrt{\kappa_{\rm e}} \eta_\mathrm{c,i}+ \sqrt{\kappa_{\rm i}}  \eta_\mathrm{c,I},\\
    \label{eq:20}
    \dot{b} =& -\dfrac{\gamma}{2}  b - i \mathcal{G} \eta_\mathrm{a} +\sqrt{\gamma}b_\mathrm{i}.  
  \end{align}
\end{subequations}
where
$\eta_\mathrm{a,i}= \cosh \xi c_\mathrm{i} + \sinh \xi {a}^\dagger_\mathrm{i}$,
$\eta_\mathrm{c, i}= \cosh \xi a_\mathrm{i} + \sinh \xi c^\dagger_\mathrm{i}$.
Eqs.~(\ref{eq:18}-\ref{eq:20}) thus show that it is possible to recast the
problem in terms of the dynamics of two operators ($\eta_\mathrm{a}$ and
$\eta_\mathrm{c}$) resulting from the action of a two-mode squeezing operator on
the original field operators, suggesting that the output modes of the field are
entangled and therefore that, potentially, nonlocal correlations are
present. For an incoming signal at the resonance frequency of either cavity, the
RWA analysis of the problem allows us to establish that in the limit of large
cooperativity ($C_-=4 G_-^2 /\kappa\gamma \gg 1$) we have that
$A_\mathrm{d}=2r_\mathrm{e}/(1-r^2)-1,\, C_\mathrm{d}=-2r_\mathrm{e}r^2/(1-r^2)-1$,
$A_\mathrm{x}=-C_\mathrm{x}=2r_\mathrm{e} r/(1-r^2)$, where $r=G_+/G_-$ and
$r_\mathrm{e}=\kappa_\mathrm{e}/\kappa$ is the ratio between the external
coupling rate to the total losses of the cavities.

Nevertheless, in our analysis, unless explicitly stated, we show the results for the full
solution of Eqs.~(\ref{eq:5}-\ref{eq:7}) (i.e. without resorting to the RWA)
and we assume that both cavities have the same environment coupling properties.

In our discussion, we will consider that, in addition to the strong coherent
tone $\alpha_{\mathrm{A}}$ and $\alpha_{\mathrm{C}}$, cavity A and cavity C are
also driven by small coherent input fields $\alpha_\mathrm{i}$ and
$\chi_\mathrm{i}$, respectively. In this scenario, the relation between input
and output fields given by Eq.~\eqref{eq:8} allows us to evaluate the response
at the output of each cavity to the fields $\alpha_\mathrm{i}$ and
$\chi_\mathrm{i}$. The correlations between $a_{\mathrm{o}}$ and
$c_{\mathrm{o}}$ introduced by the combined dynamics of the two cavities and of
the mechanical resonator represent the key ingredient for the generation of the
correlations required to violate the CHSH BI.

%% FPM more details about the detection scheme. discuss differences between MW &
%% optics 
As anticipated, the protocol that we have in mind is based on the measurement of
the field intensity at two pairs of detectors D1/E1, D2/E2 corresponding to the
photodetection scheme of the Ref. \cite{Tan:1990ea} after mixing the signals
$a_\mathrm{o}$ and $c_\mathrm{o}$ emerging from the optomechanical device with
two local oscillators (LOs).
%local oscillators (LOs). 
This detection scheme is closely related to a balanced homodyne detection setup,
in the case discussed here, however, both signals originating from the beam
splitters are recorded in order to measure the required correlations.  More
specifically, the outputs $a_{\mathrm{o}}$ and $c_{\mathrm{o}}$ of the cavities
are directed to two detectors, constituted by a beam splitter and two
photodetectors each (see Fig.~\ref{fig:1}).  At each
detector the signal field is mixed with a coherent field of a
%coherent local oscillator 
LO $\beta_{1,2}$ by a 50:50
beam splitter. The signals originating form the beam splitters are then measured
at the photodetectors D1, E1, D2, and E2. In order to evaluate the correlations
needed for the verification of the violation of the CHSH inequality, we define
the correlations pairs $D1/E1$ and $D2/E2$ for different phases of the LOs as
\begin{subequations}
\begin{align}
  &R_{+\,+}\left(\theta,\phi\right)= \Braket{d_1^\dagger d_2^\dagger  d_2 d_1},
  \label{eq:21}\\
  &R_{+\,-}\left(\theta,\phi\right)= \Braket{d_1^\dagger  e_2^\dagger e_2 d_1},
  \label{eq:22}\\
  &R_{-\,+}\left(\theta,\phi\right)= \Braket{e_1^\dagger  d_2^\dagger d_2 e_1}, 
  \label{eq:23}\\
  &R_{-\,-}\left(\theta,\phi\right)= \Braket{e_1^\dagger  e_2^\dagger  e_2 e_1},
  \label{eq:24}
\end{align}
\end{subequations}
where $d_1$/$d_2$, $e_1$/$e_2$ are the fields associated with each of pair of
photodetectors, and $\theta$ and $\phi$ represent the coherent field phases of
each LO. In the language of quantum optics, $R_{\mathrm{i\,j}}$ ($i,j=\pm$)
represent the intensity correlations among photocurrents in the 4 detectors,
e.g. $R_{+\,-}$ measures correlations between the photocurrent in $\mathrm{D}_1$
and the one in $\mathrm{E}_2$. The setup we are discussing here is analogous to
the more conventional polarization experiments
\cite{Freedman:1972ka,Fry:1976hv,Aspect:1982ja,Aspect:1982br,Weihs:1998cc,
  Giustina:2013jsa,Christensen:2013dn, Giustina:2015fc, Shalm:2015cw}: in these
experiments each channel D1/E1, D2/E2 is selected by adjusting the angle of a
polarizer at each detection branch. The parallel with the polarization
experiments, is represented by the fact that, by changing the phase of the LO,
we are selecting the detection channel, essentially performing a quadrature
measurement of the output fields originating from of the optomechanical system,
since it is possible to relate
$R_{\mathrm{i\,j}}$ in Eqs.~(\ref{eq:21}-\ref{eq:24}) to the quadratures
$X_a(\theta)=\left(a_\mathrm{o}^\dagger e^{i\theta}+a_\mathrm{o} e^{-i\theta}\right)/\sqrt{2}$
and
$X_c(\phi)=\left(c_\mathrm{o}^\dagger e^{i\phi}+c_\mathrm{o} e^{-i\phi}\right)/\sqrt{2}$
of the output fields given in Eqs. (\ref{eq:8},\ref{eq:9}). More specifically,
--focusing, for instance, on the lhs detector in Fig.~\ref{fig:1}-- we can write
the fields $d_1$ and $e_1$ as the result of the mixing between the LO field
$b_{\mathrm{LO1}}$ and $a_\mathrm{o}$, the output field of cavity $\mathrm{A}$ as 
\begin{subequations}
\begin{align}
   d_1= \sqrt{\eta_1} a_\mathrm{o} + i \sqrt{1-\eta_1} b_{\mathrm{LO1}},\label{eq:25} \\
   e_1= \sqrt{\eta_1} b_{\mathrm{LO1}} + i \sqrt{1-\eta_1} a_\mathrm{o},\label{eq:26} 
\end{align}
\end{subequations}
where $\eta_1$ is the transmissivity of the beam splitter associated with the
lhs detector of Fig.~\ref{fig:1}. Therefore, as discussed more in detail in
Appendix \ref{CH}, we can express the correlators in Eqs.~(\ref{eq:21}-\ref{eq:24})
in terms of $a_\mathrm{o}$ and $c_\mathrm{o}$.

Regardless of the physical implementation, either in the optical of the
microwave frequency range, the original formulation of the CHSH
inequality is given by the following relation
\begin{align}
  \left|S \right|=\left| E\left(\theta_1,\phi_1\right)+E\left(\theta_2,\phi_2\right)+E\left(\theta_1,\phi_2\right)-E\left(\theta_2,\phi_1\right)\right|\leq 2,
  \label{eq:27}
\end{align}
where, in our case, we have
\begin{align}
   E\left(\theta,\phi\right)=\frac{R_{+\,+}+R_{-\,-}-R_{-\,+}-R_{+\,-}}{R_{+\,+}+R_{-\,-}+R_{-\,+}+R_{+\,-}}.
   \label{eq:28}
 \end{align}
In terms of correlations of the original optomechanical fields $a_\mathrm{o}$
and $c_\mathrm{o}$, Eq.~\eqref{eq:28} can be written as
\begin{align}
  E= C \cos \left[\bar{\theta} -\bar{\phi}  \right] +  D \cos \left[\bar{\theta}  +\bar{\phi}  \right],
  \label{eq:29}
\end{align}
where
\begin{align*}
C=2|\langle a^\dagger_{\mathrm{o}} c_{\mathrm{o}}\rangle|/Z \\
D=2|\langle a_{\mathrm{o}} c_{\mathrm{o}}\rangle |/Z
\end{align*}
  with
$Z = 2\sqrt{\braket{a^\dagger_{\mathrm{o}} c^\dagger_{\mathrm{o}} c_{\mathrm{o}} a_{\mathrm{o}}}}+ \langle a^\dagger_{\mathrm{o}} a_{\mathrm{o}}+c^\dagger_{\mathrm{o}} c_{\mathrm{o}}\rangle$
and we have absorbed the phases of $\braket{a_{\mathrm{o}} c_{\mathrm{o}}}$ and
$\braket{a^\dagger _{\mathrm{o}} c_{\mathrm{o}}}$ in the definitions of
$\bar{\theta}$ and $\bar{\phi}$, and
$|\beta_1|=|\beta_2|=|\beta|= \braket{a^\dagger_{\mathrm{o}} c^\dagger_{\mathrm{o}} c_{\mathrm{o}} a_{\mathrm{o}}}^{1/4}$. It
can be shown that the latter condition maximizes the violation of the inequality
given in Eq.~(\ref{eq:27}) -- see Appendix~\ref{CH}.
% Moreover, the violation of above inequality is maximized for
%$ \left|\beta\right|^4 =\braket{a^\dagger_{\mathrm{o}} c^\dagger_{\mathrm{o}} c_{\mathrm{o}} a_{\mathrm{o}}}$ (see Appendix \ref{CH}).  
%(see also SM\cite{supp}).

The maxima of $S$ occur when $\bar{\theta}$ =0, $\bar{\phi}=-\zeta$,
$\bar{\theta}' =-\pi/2$ and $\bar{\phi}'=\zeta$ and with a maximum value is
given by
\begin{align}
S=2\sqrt{2} \sqrt{C^2+D^2} \sin(\zeta-\zeta_0),
\end{align}
where $\tan(\zeta_{0})=(C+D)/(C-D)$. It is clear that the CHSH
inequality given in Eq.~\eqref{eq:27}, can be translated into the condition \cite{Tan:1990ea}
\begin{align}
  \mathcal{F} =C^2 + D^2<\dfrac{1}{2}.
    \label{eq:30}
\end{align}

The BI test in the optomechanical setting described by Eq.~\eqref{eq:30} can be
straightforwardly evaluated considering the definitions of $C$ and $D$, and the
input-output relations given by Eqs.~(\ref{eq:2}~-~\ref{eq:4}).
%%%%%%
\begin{figure}[htb]
  \centering
  \includegraphics[width=\linewidth]{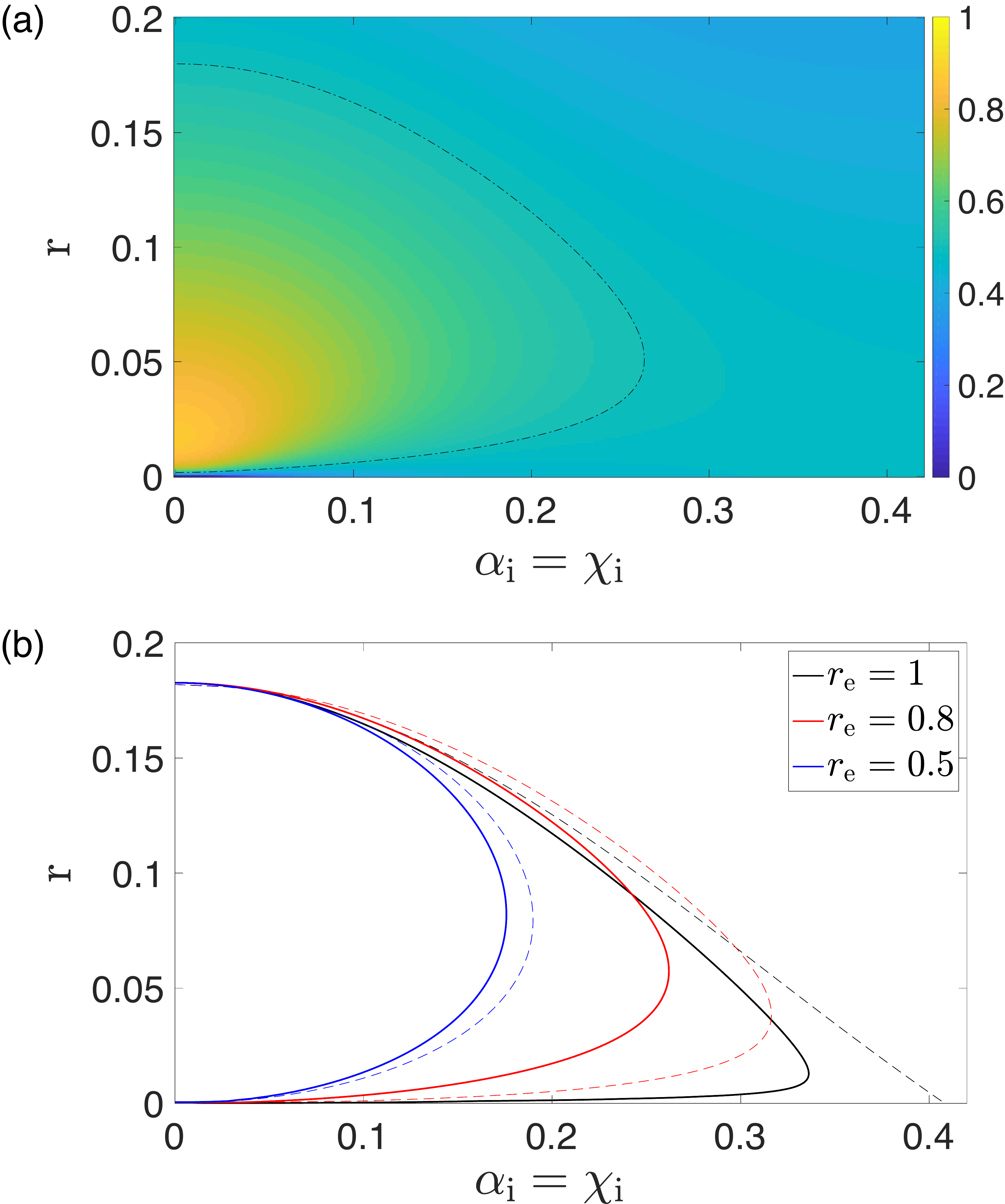}
  \caption{\textbf{a.} Value of $\mathcal{F}$ as a function of
    $\alpha_\mathrm{i}$ and $r=G_+/G_-$. Parameters:
    $\kappa_\mathrm{a}=\kappa_\mathrm{c}=\kappa = 0.1$,
    $\kappa_{\rm e }= 0.9 \kappa$, $\gamma=1 \times 10^{-5}$ -- all energies
    expressed in units of $\omega_\mathrm{m}$ ($\hbar=1$) throughout the
    manuscript. The dashed curve corresponds to the exact boundary region
    $\mathcal{F}=1/2$ , as determined from the solution of
    Eqs.~(\ref{eq:2}~-~\ref{eq:4}) \textbf{b.} Boundary $\mathcal{F} =1/2$ for
    different values of $r_{\rm e }=\kappa_{e}/\kappa$. Smaller regions are
    associated with smaller values of $r_\mathrm{e}$.  For
    $\alpha_{\rm i} \simeq 0.2$ it is possible to observe a crossing between
    boundary regions for different values of $r_\mathrm{e}$, hinting a
    nontrivial relation between entanglement and violation of the CHSH BI (see
    text). The solid line corresponds to the exact boundary as in \textrm{a.},
    the dashed line correspond to the expression given in Eq.~\eqref{eq:31}.}
  \label{fig:2}
\end{figure}

\section{Results and discussion}
\label{RD}

In Fig.~\ref{fig:2} we have plotted the value of $\mathcal{F}$ as a function of
the ratio between the linearized pump strengths $r=G_+/G_-$ and the coherent
inputs $\alpha_\mathrm{i}$ and $\chi_\mathrm{i}$ in the absence of noise sources
for parameters compatible with present-day experimental capabilities.  Form this
figure one can see that there is a finite parameters region for which the
inequality is violated. In the limit of large cooperativity ($C_-\gg 1$), the
maximum value of $r$ leading to a violation of the BI is obtained for
$\alpha_\mathrm{i}, \chi_\mathrm{i}\to 0$ and is given by
$\bar{r}=(15+4 \sqrt{14})^{-1/2}$.  Furthermore the maximum violation of the BI
$\mathcal{F}=1$ is attained for $\alpha_\mathrm{i},\chi_\mathrm{i}\to 0$ and
$r \to 0^+$. More specifically, for large cooperativity ($C_-\gg 1$), the value
of $\mathcal{F}$ exhibits a discontinuity at
$\alpha_{\rm in}\, (=\chi_{\rm in})=0$, $r=0$.  As expected, for $G_+=0$ $(r=0)$
modes $a_{o}$ and $c_{o}$ are not entangled and $\mathcal{F}=0$.

We note here that the $r-$dependence of the function $\mathcal{F}$ is contrasted
by the $r-$dependence of entanglement. From the definition of the parameters
$A_\mathrm{d}$ and $A_\mathrm{x}$, following Eqs.~\eqref{eq:8}, it is possible
to see that, since the squeezing parameter
$z=\arctanh \left[A_\mathrm{x}/A_\mathrm{d}\right] \to \infty$ for $r \to 1^-$,
one obtains an infinitely squeezed state in this regime. This seemingly
contradictory conclusion, analogous to the one derived in
\cite{Tan:1990ea,Tan:1991cg}, is however corroborated by observing that, for
mixed states, the relation between entanglement and nonlocality exhibits aspects
that are still not fully understood \cite{Brunner:2014kr}: in particular it can
be shown that maximally entangled states ($r \to 1^-$, in our case) do not
necessarily violate locality constraints, which, conversely, can be violated by
less entangled states \cite{Vidick:2011bk,Junge:2011fv,Vallone:2014id}. In our
setup, this complex interplay between entanglement and nonlocality is further
exemplified by the crossing between the $\mathcal{F}=1/2$ boundary
regions for different values of $r_\mathrm{e}$: as it is possible to see in
Fig. \ref{fig:2}, for intermediate values of the coherent drive
($\alpha_\mathrm{i}\simeq 0.1-0.2$ in this case), larger values of
$r_\mathrm{e}$ lead to a reduction of the value of  $r$ for which the violation
is observed. 

% $r\to 0^+$, $\mathcal{F}=1/2$ the entanglement is maximal.  This seemingly
% counterintuitive behaviour can be understood considering that, while $G_{+}$
% provides the entanglement between cavity A and the mechanics, in order to
% generate a faithful mapping of the mechanical mode to cavity C, a (strong) $G_-$
% tone is required.
It is clear that a violation of the CHSH inequality is possible only for small
values of the input fields $\alpha_\mathrm{i}$ and $\chi_\mathrm{i}$, and for
small values of $r$ implying $|A_{\rm d}|=|C_\mathrm{d}|\approx 1$ and
$|A_{\rm x}|=|C_{\rm x}|\ll 1$.  Therefore, in spite of the fact that the setup
proposed here has been used for nearly quantum-limited amplification
\cite{OckeloenKorppi:2016ke}, the requirements for the observation of the
violation of the BI dictate that
$\braket{a^\dagger_\mathrm{o} a_\mathrm{o}}\simeq |A_\mathrm{d}|^2 \braket{a^\dagger_\mathrm{i} a_\mathrm{i}}\approx 0.1$
and
$\braket{c^\dagger_\mathrm{o} c_\mathrm{o}}\simeq |C_\mathrm{d}|^2 \braket{c^\dagger_\mathrm{i} c_\mathrm{i}}\approx 0.1$.
This condition combines the concomitant requirements that the value of
$\mathcal{F}$ and the output signals have to be maximized. In order to gain
better insight on the range of physical parameters for which the BI inequality
is violated, we can establish an approximate analytical expression for the
maximum value of $\alpha_\mathrm{i}$ violating the inequality as
\begin{align}
  \alpha_{\rm i}=\sqrt{r_\mathrm{e} \bar{r} \left(1- 4\, \bar{r} -6 \,\bar{r}^2- 1 2\, \bar{r}^3 \right)
                       /\left(\mathcal{K}_0\, \bar{r}^2+\mathcal{K}_1 \, \bar{r}  +\mathcal{K}_2 \right)}\, ,
 \label{eq:31}
\end{align}
where $\mathcal{K}_0=28 r^2_{\rm e}$,
$\mathcal{K}_1=2 (1-2 r_{\rm e} + 4 r^2_{\rm e})$ and
$\mathcal{K}_2=2(1-r_{\rm e})^2$.
Eq.~\eqref{eq:31} is obtained as a second-order expansion of $\mathcal{F}$ in the input field
intensity $\alpha_\mathrm{i}^2$ evaluated here for the RWA solution of the
problem.
\begin{figure}[htb]
  \centering
  \includegraphics[width=\linewidth]{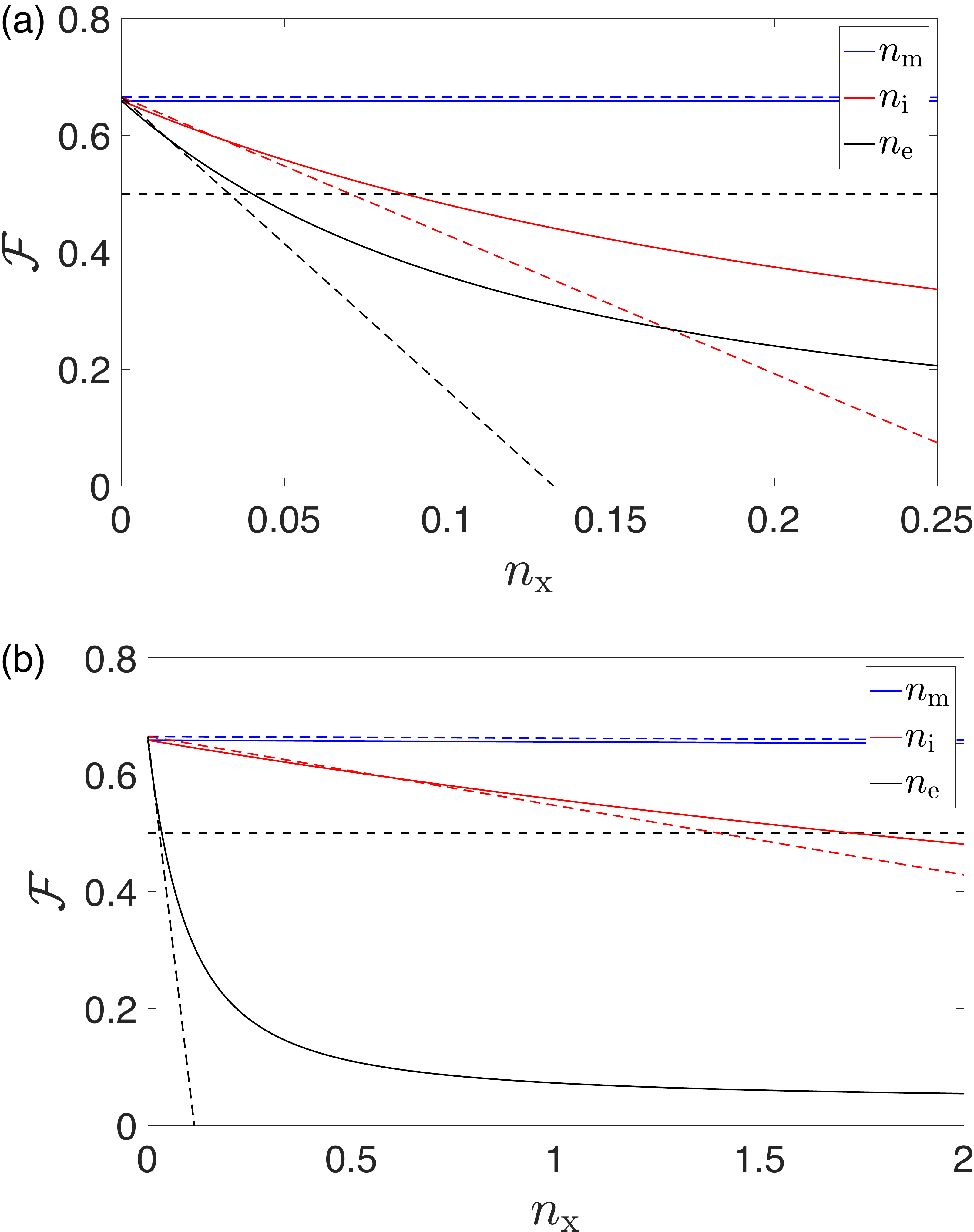}
  \caption{Dependence of the value of $\mathcal{F}$ (for
    $r \to r_\mathrm{{opt}} $ and $\alpha_\mathrm{i}$, $\chi_{\rm i} \to 0$) on
    the thermal population baths associated with the mechanical noise
    ($\bar{n}_\mathrm{m}$, blue --flattest-- curve), internal noise
    ($\bar{n}_\mathrm{i}$, red --intermediate-- curve) and external noise
    ($\bar{n}_\mathrm{e}$, black --steepest-- curve). Solid lines correspond to
    the exact solution from the equations of motion with each noise source
    considered independently. Dashed lines are the approximations given in
    Eq.~\eqref{eq:16} and Eqs.(\ref{eq:33}-\ref{eq:36}).  Values of
    $\mathcal{F}$ above the horizontal dashed line at $\mathcal{F}=1/2$
    correspond to the violation of the BI.  Parameters: \textbf{a.}
    $\kappa=0.01$, $\kappa_e=0.9\, \kappa$, $G_-=0.2$, $\gamma=1 \cdot 10^{-5}$.
    \textbf{b.} $\kappa_e=0.99\, \kappa$, all other parameters as in
    \textbf{a.}.}
  \label{fig:3}
\end{figure}

So far, the discussion has focused on the ideal situation for which the effect
of noise is negligible. In the following, we address the role played by the
different environmental noise sources. In particular, we take into account the
presence of a thermal environment for the mechanical resonator
($\bar{n}_{\rm m}$, ``mechanical noise''), for the two resonant cavities
($\bar{n}_\mathrm{i}$, ``internal noise'') and to the noise associated with the
coupling of the two resonant cavities to the input and output ports
($\bar{n}_\mathrm{e}$, ``external noise''). Without loss of generality, in
Eq.~\eqref{eq:32} we have assumed that the noise temperature for the two
cavities is equal and that all noise sources are independent.  If we consider
the effect of the noise on $\mathcal{F}$ to the first order, we can write
\begin{align}
\mathcal{F}=\mathcal{F}_0 - \mathcal{F}_{\rm m}  \bar{n}_{\rm m} - \mathcal{F}_{\rm e} \bar{n}_{\rm e} - \mathcal{F}_{\rm i} \bar{n}_{\rm i},  
\label{eq:32}
\end{align}
where $\mathcal{F}_0$ is the quantity previously considered for the violation of
the BI, the second term represents the contribution associated with the
mechanical noise, and the third (fourth) term describes the external (internal)
noise contribution due to the thermal environment associated with the cavity
modes. The sensitivity of the BI violation to the noise terms is encoded in the
coefficients $\mathcal{F}_{\rm e}$, $\mathcal{F}_{\rm i} $ and
$\mathcal{F}_{\rm m}$: the larger the coefficients, the more each noise term
contributes to the reduction of the value of $\mathcal{F}$ and, therefore, to
the reduction of the region for which the BI is violated. An approximate
expression for the factors appearing in Eq.~\eqref{eq:32} can be obtained
expanding the RWA approximation for $\mathcal{F}_\mathrm{0}$,
$\mathcal{F}_\mathrm{m}$, $\mathcal{F}_\mathrm{e}$, $\mathcal{F}_\mathrm{i}$, to
the lowest order in $1/C_-$
\begin{subequations}
  \begin{align}
    \mathcal{F}_\mathrm{0}&= \left(2 r -1\right)^2 +4 r^2 \label{eq:33} \\
    \mathcal{F}_\mathrm{m}&=-2\frac{ \left(2 r -1\right)^2 +2 r^2 \left(10 r-1\right)}{C_-} \label{eq:34} \\
    \mathcal{F}_\mathrm{e}&=-\frac{\left(2 r -1\right)^2 + r^2\left(16 r-1\right)}{r}\frac{r_\mathrm{e}^2+r_\mathrm{i}^2}{r_\mathrm{e}} \label{eq:35} \\
    \mathcal{F}_\mathrm{i}&=-\frac{\left(2 r -1\right)^2  + r^2\left(16 r-1\right)}{r}r_\mathrm{i} \label{eq:36} 
  \end{align}
\end{subequations}
The portion of the noise associated with the mechanics and described in the
linear approximation by $\mathcal{F}_\mathrm{m}$ --see Eq.~\eqref{eq:34}-- can
be modified by tuning the parameter $C_-$. This dependence can be understood as
the result of a sideband cooling process operated by the drive of cavity C,
which is driven on the red sideband. In addition, $\mathcal{F}_\mathrm{i}$ can
be reduced by minimizing the contribution of internal losses --see
Fig.~\ref{fig:3}--, whereas $\mathcal{F}_\mathrm{e}$ cannot be altered significantly
and thus represents the most critical parameter.
\begin{figure}[htb]
\centering
\includegraphics[width=0.9\linewidth]{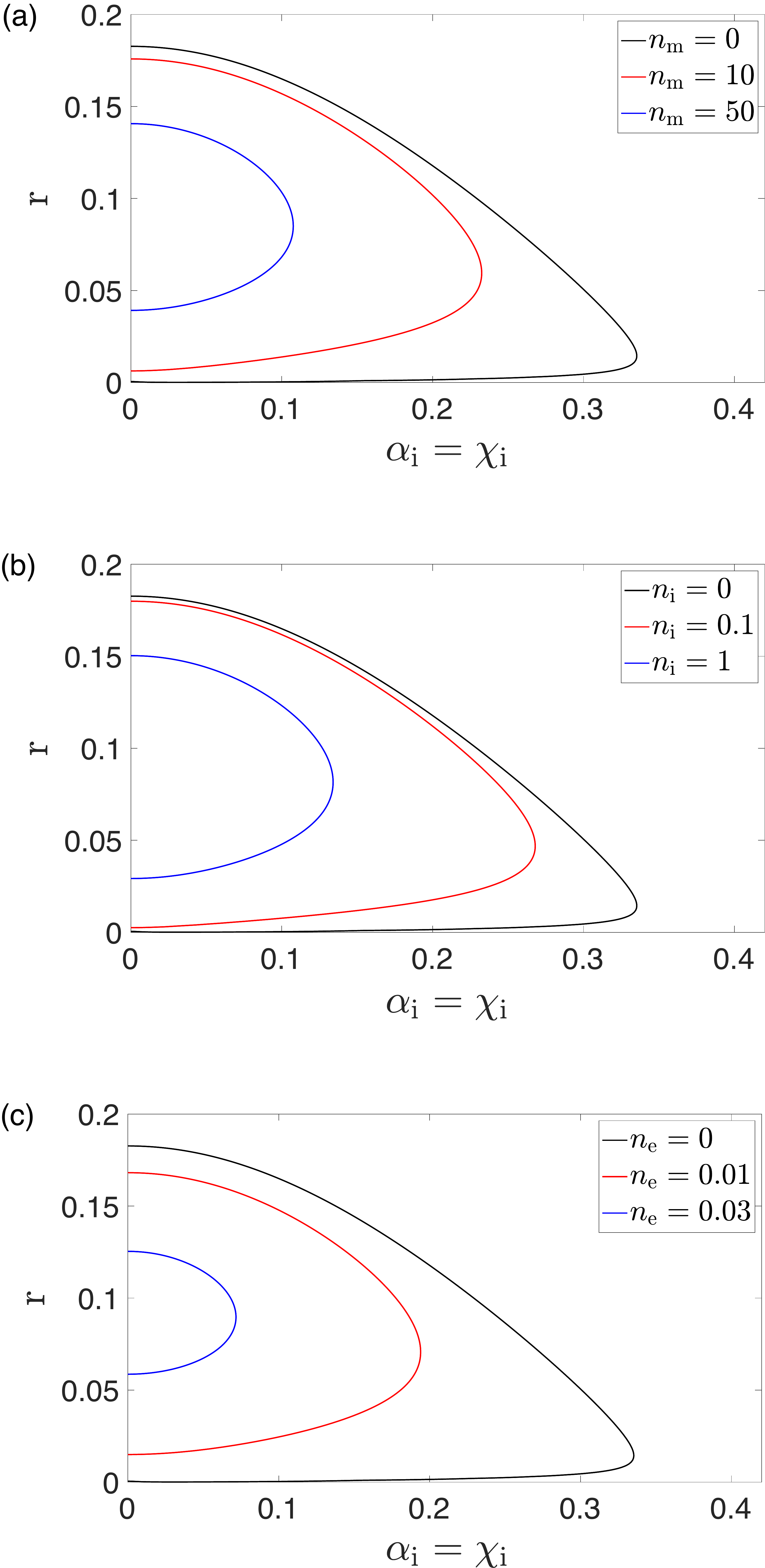}
\caption{Noise-dependence of the $\mathcal{F}=1/2$ boundary in the presence of a
  finite coherent input. Smaller regions correspond to large value of the
  noise. All parameters as in Fig.~\ref{fig:2}(b).}
\label{fig:4}
\end{figure}

This conclusion is corroborated by Fig.~\ref{fig:4}, where we have depicted the
separate effects of different noise sources on the value of $\mathcal{F}$, it is
clear that the input noise $\bar{n}_{\rm e}$ represents the most sensitive
parameter in the violation of the CHSH inequality. In this perspective, we thus
select a value of $r$ that, whilst representing a sub-optimal choice
(i.e. $\mathcal{F}<1$) for the noiseless case, allows for the largest possible
value of $n_\mathrm{e}$ and $n_\mathrm{i}$ compatible with the violation of the
BI given in Eq. \eqref{eq:30}. In the linearized regime described by
Eq. \eqref{eq:32}, in the presence of cavity (external and internal) noise only,
the relation describing the boundary for the violation of the BI can be
expressed as
\begin{align}
  \mathcal{F}_\mathrm{0}(r)+ \mathcal{F}_\mathrm{e}(r) n_\mathrm{e} +
                                     \mathcal{F}_\mathrm{i}(r) n_\mathrm{i} =\frac{1}{2}
  \label{eq:71}
\end{align}
where we have supposed that $r_\mathrm{e}$ is held fixed. From
Eqs.~(\ref{eq:33}-\ref{eq:36}) we can write Eq.~(\ref{eq:71}) as
\begin{align}
  \mathcal{F}_\mathrm{0}(r)-\frac{1}{2}+ \mathcal{F}_\mathrm{T}(r) n_\mathrm{T} =0
  \label{eq:72}
\end{align}
where
$\mathcal{F}_\mathrm{i}/r_\mathrm{i}=\mathcal{F}_\mathrm{i} r_\mathrm{e}/\left(r_\mathrm{e}^2+r_\mathrm{i}^2\right)=\mathcal{F}_\mathrm{T}$
and
$n_\mathrm{T}=\left(r_\mathrm{e}^2+r_\mathrm{i}^2\right)/r_\mathrm{e}  n_\mathrm{e}+r_\mathrm{i} n_\mathrm{i}$. From
Eq.~\eqref{eq:72}
$n_\mathrm{T}=\left[1/2-\mathcal{F}_\mathrm{0}(r)\right]/\mathcal{F}_T(r)$ can
be straightforwardly maximized yielding the optimal value for $r=r_\mathrm{opt}$.

We would like to stress however that the contribution associated with
$n_\mathrm{e}$ assumes that the baths for the cavities are uncorrelated with
each other, which represents a somewhat worst-case scenario. The potential
presence of correlated noise can be considered, from the perspective of the BI
violation, as a contribution to the input signals $\alpha_\mathrm{i}$ and
$\chi_\mathrm{i}$.

For a microwave setting, we can assume that the cavity internal and external
thermal populations are set by the base temperature of the dilution fridge
($T=\SI{7}{mK}$, $\omega_\mathrm{c}=2 \pi\, \SI{10}{\giga\hertz} $)
corresponding to $n_\mathrm{i}=n_\mathrm{e} \simeq 0.015$, whereas for an
optical setting at room temperature ($T=\SI{300}{K} $,
$\omega_\mathrm{c}=2 \pi\, \SI{500}{THz}$) we have
$n_\mathrm{i}=n_\mathrm{e} \simeq 0.02$.  While in both cases the deviation from
ideality is significant, the BI is still clearly violated both for the microwave
setting ($\mathrm{F} \simeq 0.56$, for $r_\mathrm{e} =0.9$,
$\mathrm{F} \simeq 0.58$, for $r_\mathrm{e} =0.99$) and for the optical case
($\mathcal{F} \simeq 0.59$, for $r_\mathrm{e} =0.9$, $\mathcal{F} \simeq 0.60$,
for $r_\mathrm{e} =0.99$). In Fig.~\ref{fig:3}, it is
possible to note that, for parameters compatible with microwave realizations of
the setup discussed in this article, the mechanical noise does not contribute to
the reduction of $\mathcal{F}$. This effect is closely related to the physics of
the quantum-limited amplifier discussed in Ref. \cite{OckeloenKorppi:2016ke}: in
both cases the mechanics, while mediating the interaction required to generate
the output fields, is concomitantly cooled by the pumping tones.

\section{Conclusion}\label{con}
We have discussed here a potential CHSH Bell inequality test based on a
quadrature phase coherence measurement in an optomechanical setting. We have
shown that it is possible to violate the CHSH Bell inequality in an optomechanical
setting by weakly driving two cavity/ one mechanics device. Furthermore, we have
demonstrated that, while the thermal noise associated with cavities and
mechanical degrees of freedom degrades the performances of the device proposed
here, the latter is naturally suppressed by the working principle of our device.
We hypothesize that our proposal could be implemented either in an optical or
in a circuit QED setting.
\section*{Acknowledgments}
We thank Elli Selenius, Mika Sillanp{\"a}{\"a} and Caspar F. Ockeloen-Korppi for
useful discussions.  This work was supported by the Academy of Finland (Contract
No. 275245) and the European Research Council (Grant No. 670743).

\pagebreak
\onecolumngrid
\appendix
\section{Equations of motion}
\label{DM}
We derive here the equations of motion for the 2 cavities / 1 mechanical
resonator system given in Eqs.~(\ref{eq:2},\ref{eq:4}) of the main text. In the
presence of a strong coherent tones at blue (red) sideband for cavity A (C), the
quantum Langevin equations associated with the Hamiltonian given in Eq.~\eqref{eq:1} of
the main tex can be written as
\begin{subequations}
\begin{align}
\label{eq:37}
\dot{a}=& -(i\omega_\mathrm{a}+\dfrac{\kappa_\mathrm{a}}{2}) a - i g_\mathrm{a} a (b + b^\dagger) +
% a_\mathrm{in, A} e^{-i\omega_\mathrm{d,A}t}+
\sqrt{\kappa_\mathrm{e,a}}\, a_\mathrm{i} + \sqrt{\kappa_\mathrm{i,a}}\, a_\mathrm{I},\\
%%%%%%%% 
\label{eq:38}
\dot{c}=& -(i\omega_\mathrm{c}+\dfrac{\kappa_\mathrm{c}}{2}) c - i g_\mathrm{c} c (b + b^\dagger) +\sqrt{\kappa_\mathrm{e,c}} \, c_\mathrm{i} + \sqrt{\kappa_\mathrm{i,c}}\, c_\mathrm{I},\\
%+\alpha_\mathrm{in, C} e^{-i\omega_\mathrm{d,C}t}+\sqrt{\kappa_\mathrm{c}}c_\mathrm{i},\\
  \label{eq:39}
\dot{b}=& -(i\omega_\mathrm{m}+\dfrac{\gamma}{2} )b - i g_\mathrm{a} a^\dagger a - i g_\mathrm{c} c^\dagger c+\sqrt{\gamma}\, b_\mathrm{i},
\end{align}
\end{subequations}
where $\kappa_\mathrm{a}=\kappa_{\mathrm{e,a}} + \kappa_{\mathrm{i,a}}$ is the
total cavity decay rate where $\kappa_{\mathrm{i, a}}$ and
$\kappa_{\mathrm{e, a}}$ are the internal and external cavity decay rates,
(analogous relations hold for cavity C). The fields $a_\mathrm{i}$,
$c_\mathrm{i}$, $b_\mathrm{i}$, represent the input fields driving the cavities
and the mechanical resonator, whereas $a_\mathrm{I}$  and
$c_\mathrm{I}$ describe the contributions from the internal
noise for cavity A and cavity C, respectively.
%$a_\mathrm{i}$  and $c_\mathrm{i}$ are the external drive amplitudes for two cavities A and C.
In the main text we consider the case of a strong drive for both cavities ( with
amplitudes $\alpha_{\mathrm{in, A}}$ and $\alpha_{\mathrm{in, C}}$, at
frequencies $\omega_{\mathrm{d,A}}$ and $\omega_{\mathrm{d,C}}$,
respectively). In this case, the quantum Langevin equations given in Eqs.~(\ref{eq:37}-\ref{eq:39}) can
be linearized around the the cavity fields induced by the pump tones, leading to
the following expression for the steady state for the cavity fields

\begin{subequations}
  \begin{align}
    \bar{\alpha}_{\rm A}=\dfrac{\alpha_{\rm in,A}}{\dfrac{\kappa_{\rm a}}{2}+i( \omega_{\rm a}
                                - g_{\rm a}\alpha_{\rm A} ( b_{\rm s}+b^*_{\rm s}))} e^{-i\omega_\mathrm{d,A}t}=\alpha_{\rm A}  e^{-i\omega_\mathrm{d,A}t}, \\
    \bar{\alpha}_{\rm C}=\dfrac{\alpha_{\rm in,C}}{\dfrac{\kappa_{\rm c}}{2}+i( \omega_{\rm c}
                                - g_{\rm c}\alpha_{\rm C}( b_{\rm s}+b^*_{\rm s}))} e^{-i\omega_\mathrm{d,C}t}=\alpha_{\rm C}  e^{-i\omega_\mathrm{d,C}t}
  \end{align}
\end{subequations}
while the equations for the fluctuations around the steady-state values are given by
%optical and mechanical variables $(\bar{\alpha}_{\rm A}, \bar{\alpha}_{\rm C}$). The steady

 %and $\bar{\alpha}_{\rm B})$, neglecting nonlinear and c-number terms we get 
\begin{subequations}
\begin{align}
\label{eq:40}
  \dot{a}=& -(i\omega_\mathrm{a}+\dfrac{\kappa_\mathrm{a}}{2}) a
                 - i g_\mathrm{a} \bar{\alpha}_{\rm A} (b + b^\dagger) +\sqrt{\kappa_\mathrm{e,a}}\, a_\mathrm{i} + \sqrt{\kappa_\mathrm{i,a}}\, a_\mathrm{I},\\
%%%%%%%%%%%%%%%%%
%%%%%%%%
\label{eq:41}
\dot{c}=& -(i\omega_\mathrm{c}+\dfrac{\kappa_\mathrm{c}}{2}) c - i g_\mathrm{c} \bar{\alpha}_{\rm C} (b + b^\dagger) 
+\sqrt{\kappa_\mathrm{e,c}}\, c_\mathrm{i} + \sqrt{\kappa_\mathrm{i,c}}\, c_\mathrm{I},\\
%%%%%%%%%%%%
\label{eq:42}
\dot{b}=& -(i\omega_\mathrm{m}+\dfrac{\gamma}{2} )b - i g_\mathrm{a} \bar{\alpha}_{\rm A} (a + a^\dagger) - i g_\mathrm{c} \bar{\alpha}_{\rm C} (c+ c^\dagger) +\sqrt{\gamma}b_\mathrm{i}.
\end{align}
\end{subequations}
%where $\bar{\alpha}_{\rm A}= \alpha_\mathrm{A} e^{-i\omega_\mathrm{d,A}t}$ and $ \bar{\alpha}_{\rm C}= \alpha_\mathrm{C} e^{-i\omega_\mathrm{d,C}t}$.  
%Here, we consider that the cavity A is driven at blue ( $\omega_\mathrm{d,A}=\omega_\mathrm{a} + \omega_\mathrm{m}$) and cavity B is driven at red ( $%\omega_\mathrm{d,C}=\omega_\mathrm{c}-\omega_\mathrm{m}$) mechanical sidebands.
Moving to a frame rotating at $(\omega_{\rm d, a}, \omega_{\rm d, c}$ and
$\omega_{\rm m}$ for cavity A, cavity C and mechanics respectively), by
substituting the values of $\bar{\alpha}_{\rm A}$ and $\bar{\alpha}_{\rm C}$ in
Eqs.~(\ref{eq:40}-\ref{eq:42}), the corresponding linearized quantum Langevin
equations for the fluctuations around the stationary values induced by the pumps
(Eqs.~(\ref{eq:2},\ref{eq:4}) of the main text), are
\begin{subequations}
  \begin{align}
    \label{eq:43}
    \dot{a} =& \left(-i \Delta_\mathrm{a} -\dfrac{\kappa_\mathrm{a}}{2}\right)  a - i G_+  \left(b^\dagger + b \right)+\sqrt{\kappa_\mathrm{e,a}}\,a_{\mathrm{i}}+ \sqrt{\kappa_\mathrm{i,a}}\,a_{\mathrm{I}},\\
    \label{eq:44}
    \dot{c} =& \left(-i \Delta_\mathrm{c}-\dfrac{\kappa_\mathrm{c}}{2}\right)  c -i G_-  \left(b^\dagger + b \right) + \sqrt{\kappa_\mathrm{e,c}}\,c_{\mathrm{i}}+ \sqrt{\kappa_\mathrm{i,c}}\,c_{\mathrm{I}}, \\
    \label{eq:45}  
    \dot{b} =& \left(-i \omega_\mathrm{m} -\dfrac{\gamma}{2}\right) b - i G_+ \left(a^\dagger+ a\right) - i G_- \left(c^\dagger + c\right) + \sqrt{\gamma}\, b_{\mathrm{i}},
  \end{align}
\end{subequations}
where $G_+=g_\mathrm{a} \alpha_\mathrm{A}$ and
$G_-=g_\mathrm{c} \alpha_\mathrm{C}$ are the effective linearized couplings
(without loss of generality, hereafter we assume that
$\kappa_{\rm a}=\kappa_{\rm c}=\kappa$).

\section{Input/output equations in the rotating-wave approximation}
\label{IOeq}
While the coefficients $A_\mathrm{d}$, $A_\mathrm{x}$, $C_\mathrm{d}$,
$C_\mathrm{x}$ --and therefore the condition expressing the violation of the
BI--given in Eqs.~(\ref{eq:8},\ref{eq:9}) of the main text can be obtained
without resorting to RWA, in order to outline the essential physical process
behind our proposal, we determine here the explicit analytical expression for
these coefficients within the RWA.

\begin{figure}[htb]
\centering
\includegraphics[width=0.6\linewidth]{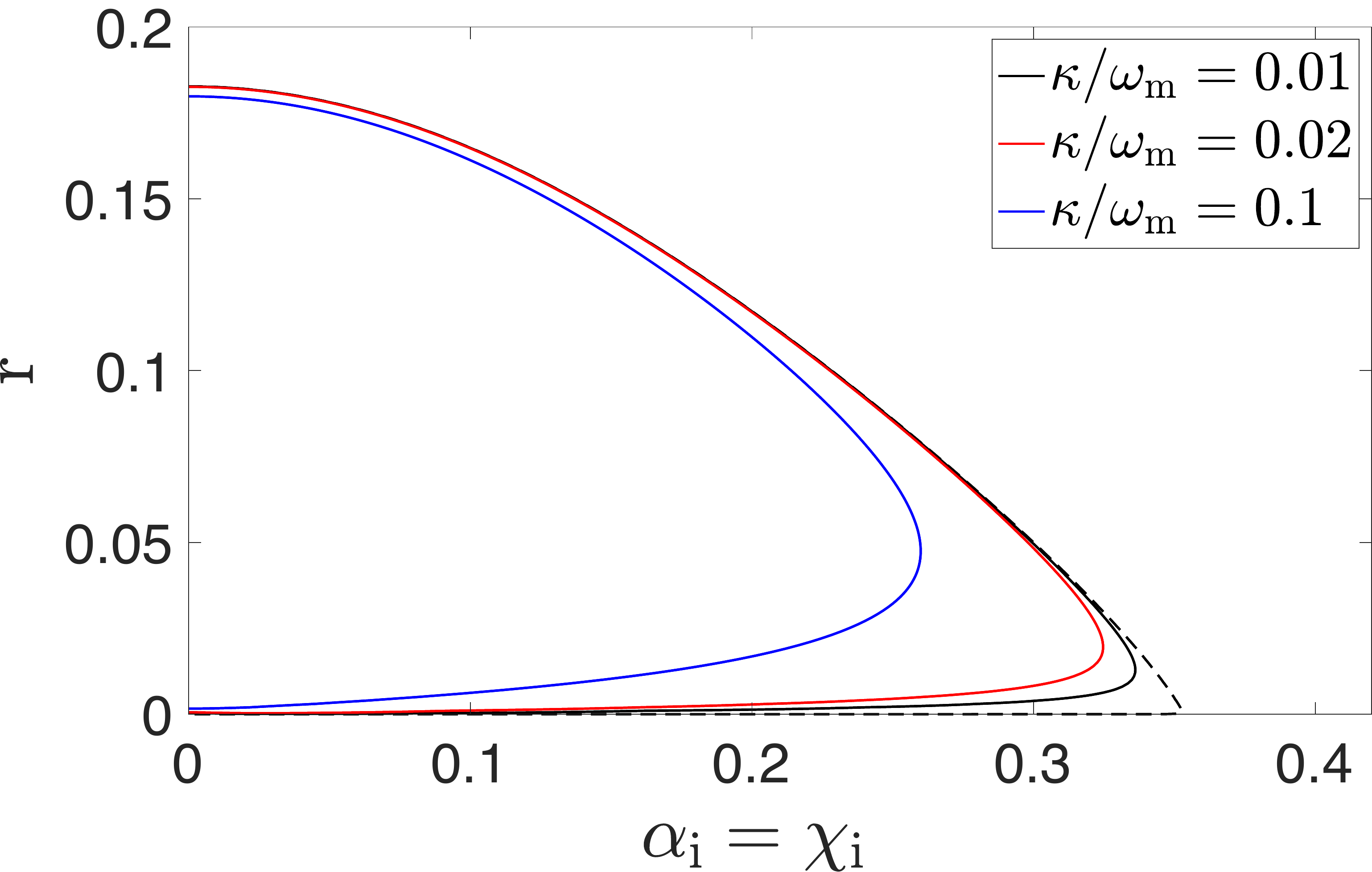}
\caption{Comparison between the value of $\mathcal{F}$ calculated from the full
  solution of the equations of motion~(full
  lines,$\kappa=0.01,\,0.02,\,0.1$, larger values correspond to
  smaller regions for which $\mathcal{F}>1/2$), with the solution obtained in
  the rotating-wave approximation (dashed line).}
\label{fig:5}
\end{figure}

In Fig.~\ref{fig:5}, it is possible to see how the validity of the RWA in the
determination of the BI violation relies on the condition
$\omega_\mathrm{m} \ll 1$ (good cavity limit) as it is usually the case in the
description of sideband pumping setups in optomechanics.
In order to derive the expression of the I/O coefficients $A_\mathrm{d}$,
$A_\mathrm{x}$, $C_\mathrm{d}$, $C_\mathrm{x}$ within the RWA, we define a
Bogolyubov unitary transformation of the optical modes operator as
\begin{subequations}
  \begin{align}
    \label{eq:57}
    \eta_\mathrm{a}= \cosh \xi \, c + \sinh \xi \,a^\dagger ,\\
    \label{eq:58}
    \eta_\mathrm{c}= \cosh \xi \, a + \sinh \xi \,c^\dagger, 
  \end{align}
\end{subequations}
where $\cosh \xi=G_-/\mathcal{G}$, $\sinh \xi=G_+/\mathcal{G}$ with
$\mathcal{G}=\sqrt{G^2_- - G^2_+}$ and rewrite Eq.~(\ref{eq:43}-\ref{eq:45}) in
terms of the Bogolyubov modes $\eta_\mathrm{a}$ and $ \eta_\mathrm{c}$ as
\begin{subequations}
  \begin{align}
    \label{eq:59}
    \dot{\eta}_\mathrm{a} =& -\dfrac{\kappa}{2} \eta_\mathrm{a} - i \mathcal{G} b + \sqrt{\kappa_{\rm e}}  \eta_\mathrm{a,i}+ \sqrt{\kappa_{\rm i}}  \eta_\mathrm{a,I},\\
    \label{eq:60}
    \dot{\eta}_\mathrm{c} =& -\dfrac{\kappa}{2} \eta_\mathrm{c} + \sqrt{\kappa_{\rm e}} \eta_\mathrm{c,i}+ \sqrt{\kappa_{\rm i}}  \eta_\mathrm{c,I},\\
    \label{eq:61}
    \dot{b} =& -\dfrac{\gamma}{2}  b - i \mathcal{G} \eta_\mathrm{a} +\sqrt{\gamma}b_\mathrm{i}.  
  \end{align}
\end{subequations}
where
$\eta_\mathrm{a,i}= \cosh \xi c_\mathrm{i} + \sinh \xi {a}^\dagger_\mathrm{i}$,
$\eta_\mathrm{c, i}= \cosh \xi a_\mathrm{i} + \sinh \xi c^\dagger_\mathrm{i}$.
We then transform the quantum Langevin equations of the two Bogolyubov
modes $\eta_\mathrm{a}$ and $\eta_\mathrm{c}$ to Fourier domain
% using $f(\omega)= \int^\infty_{-\infty} f(t) e^{-i\omega t} dt$:  
\begin{subequations}
  \begin{align}
    \label{eq:62}
    \eta_\mathrm{a}  =&\dfrac{\chi_\mathrm{a}}{1+\chi_\mathrm{m} \chi_\mathrm{a} \mathcal{G}^2}( \sqrt{\kappa_{\rm e}} \eta_\mathrm{a,i}+\sqrt{\kappa_{\rm i}} \eta_\mathrm{a,I} ) - i\dfrac{\chi_\mathrm{m} \chi_a\mathcal{G}}{1+\chi_\mathrm{m} \chi_a\mathcal{G}^2}\sqrt{\gamma}b_\mathrm{i},\\
%%%%%%%%
    \label{eq:63}
    \eta_\mathrm{c} =&\chi_\mathrm{a}  (\sqrt{\kappa_{\rm e}} \eta_\mathrm{c, i}+\sqrt{\kappa_{\rm i}} \eta_\mathrm{c, I}),
  \end{align}
\end{subequations}
where $\chi_\mathrm{m}=\left(\dfrac{\gamma}{2} -i \omega\right)^{-1}$ and
$\chi_\mathrm{a}=\left(\dfrac{\kappa}{2} -i \omega\right)^{-1}$.  Since, according to
the input-output theory \cite{Walls:2008em}, the operator for the output field
is related to the cavity and to the input noise operator by the relation
$a_\mathrm{o}=\sqrt{\kappa_\mathrm{e}} a - a_\mathrm{i}$ and
$c_\mathrm{o}=\sqrt{\kappa_\mathrm{e}} c - c_{\mathrm{i}}$ by using the
transformation
$a= \cosh \xi \eta_\mathrm{c} - \sinh \xi \eta^\dagger_\mathrm{a}$ and
$c= \cosh \xi \eta_\mathrm{a} - \sinh \xi \eta^\dagger_\mathrm{c}$, the outputs
of the two cavity modes can be written as
\begin{align}
\label{eq:64}
a_\text{o}=& 
 (\kappa_\text{e} \mathcal{A}_{\rm aa}-1) a_\text{i} + \kappa_\text{e} \mathcal{A}_{\rm ac} {c^\dagger_\text{i}}  +\sqrt{\kappa_\text{i}\kappa_\text{e}} \mathcal{A}_{\rm aa} a_\text{I}  + \sqrt{\kappa_\text{i}\kappa_\text{e}} \mathcal{A}_{\rm ac} c^\dagger_\text{I}
+ i \sqrt{\gamma \kappa_\text{e}}\frac{G_+}{(\chi_{\rm a} \chi_{\rm m})^{-1} + \mathcal{G}^2}b^\dagger_\text{i}, \\ 
%%%%%%%%%%%%%%%%%%%%%%%%
c_\text{o}=&  (\kappa_\text{e} \mathcal{A}_{\rm cc} - 1) c_\text{i} + \kappa_\text{e} \mathcal{A}_{\rm ca} a^\dagger_\text{i} + \sqrt{\kappa_\text{i}\kappa_\text{e}} \mathcal{A}_{\rm cc} c_\text{I}
+ \sqrt{\kappa_\text{i}\kappa_\text{e}} \mathcal{A}_{\rm ca} a^\dagger_\text{I}- i \sqrt{\gamma \kappa_{\rm e}}\frac{G_-}{(\chi_{\rm a} \chi_{\rm m})^{-1} + \mathcal{G}^2}b_\text{i},
\label{eq:65}
\end{align}
where
\begin{subequations}
\label{eq:66}
\begin{align}
\mathcal{A}_{ \rm aa} =& \chi_{\rm a} \cosh^2\xi - \chi^{\rm e}_{\rm a} \sinh^2\xi,\qquad \mathcal{A}_{\rm cc} = \chi^e_{\rm a} \cosh^2\xi - \chi_{\rm a} \sinh^2\xi,  \\
\mathcal{A}_{\rm ac} =& (\chi_{\rm a }- \chi^e_{\rm a}) \cosh\xi \sinh\xi,\qquad \mathcal{A}_{\rm ca} = (\chi^{\rm e}_{\rm a} - \chi_{\rm a}) \cosh\xi \sinh\xi,
\end{align}
\end{subequations}
and
$\chi_\mathrm{a}^{\rm e} =\chi_{\rm a} \left( 1 + \mathcal{G}^2 \chi_{\rm a} \chi_{\rm m}\right)^{-1}$
represents the effective cavity response in presence of the two-tone
optomechanical drive.  It is possible to write Eq.~(\ref{eq:64}-\ref{eq:65}) in
more compact form as given in Eqs.~(\ref{eq:8},\ref{eq:9}) of the main text as
\begin{subequations}
\begin{align}
a_\text{o} =& \Ad   a_\text{i} + \Ax  c^\dagger_\text{i} +\mathcal{N}_{\rm a}\label{eq:67} \\
%+\AxI c^\dagger_{\text{I}}+ A_\text{m} b^\dagger_\text{i}  
c_\text{o} =& \Cd   c_\mathrm{i} + \Cx a^\dagger_\mathrm{i} +\mathcal{N}_{\rm c}\label{eq:68}
%+ \CdI c_\text{I} + \CxI  a^\dagger_\text{I} + C_\text{m} b_\text{i} 
\end{align}
\end{subequations}
where 
\begin{align}
\mathcal{N}_{\rm a}=& \AdI a_\mathrm{I}+\AxI c^\dagger_{\text{I}}+ A_\text{m} b^\dagger_{\rm i}, \nonumber \\
\mathcal{N}_{\rm c}=& \CdI c_\text{I} + \CxI  a^\dagger_\text{I} + C_\text{m} b_{\rm i}   \nonumber
\end{align}
represent the operators associated with the mechanical and cavity internal
noise. Furthermore, the coefficients relating input and noise operators to the
output are given by
\renewcommand{\arraystretch}{2}
\begin{align*}
  \begin{array}{lll}
       A_\text{d}=\kappa_{\rm e} \mathcal{A}_{\rm aa} - 1,\qquad
    & C_\text{d}=\kappa_{\rm e} \mathcal{A}_{\rm cc} - 1,\qquad
    & A_\text{m}= +i\sqrt{\gamma \kappa_{\rm e}} G_+ \chi_{\rm a}^{\rm e}/\chi_{\rm a}\\
    %%%%%%%%%%
       A_\text{x}= \kappa_{\rm e}\mathcal{A}_{\rm ac},\qquad
    & C_\text{x}= \kappa_{\rm e}\mathcal{A}_{\rm ca}, \qquad
    & C_\text{m}=-i\sqrt{\gamma \kappa_{\rm e}} G_- \chi_{\rm a}^{\rm e} /\chi_{\rm a} \\
    %%%%%%%%%%
       \AdI= \sqrt{\kappa_\mathrm{i}\kappa_\mathrm{e}}\mathcal{A}_{\rm aa},
    & \CdI=\sqrt{\kappa_\mathrm{i}\kappa_\mathrm{e}}\mathcal{A}_{\rm cc}, \\
    %%%%%%%%%%
       \AxI=\sqrt{\kappa_\mathrm{i}\kappa_\mathrm{e}}\mathcal{A}_{\rm ac}, \qquad
    & \CxI=\sqrt{\kappa_\mathrm{i}\kappa_\mathrm{e}}\mathcal{A}_{\rm ca}. \qquad
    \end{array}
\end{align*}
 In the limit of large cooperativity $C_-=4 G^2/\kappa \gamma \gg1 $ and at the
 cavity resonance, the coefficients can be written as
 \renewcommand{\arraystretch}{3}
\begin{align*}
  \begin{array}{llll}
      \Ad=\frac{2 r_e}{1-r^2}-1,\qquad
    &\Cd=- \dfrac{2 r_e r^2}{1 - r^2}-1, \qquad
    & \Ax=\frac{2 r r_e}{1-r^2}=-\Cx, \qquad
    &\Am=-i\frac{2  r \sqrt{r_e}}{\sqrt{C_-} \left(1-r^2\right)}=r \Cm, \\
       \AdI=\frac{2 \sqrt{r_e r_i}}{1-r^2},
    & \CdI=-r^2 \AdI,
    & \AxI=\frac{2 r \sqrt{r_e r_i}}{1-r^2}=-\CxI,   
   \end{array}
\end{align*}
where $r=G_+/G_-$, $r_{\rm e}=\kappa_\mathrm{e}/\kappa$ and $r_{\rm i}=\kappa_\mathrm{i}/\kappa$.

\section{CHSH violation}
\label{CH}
We derive here the relation between the usual condition for the violation of
CHSH inequality expressed by Eq.~\eqref{eq:27}, and Eq.~\eqref{eq:30} of the
main text. To this end, we evaluate the quantity defined in Eq.~(\ref{eq:21}-\ref{eq:24}) of the main
text in terms of the output correlators of the optomechanical system. For beam
splitters of transmissivity given by $\eta_{1}$ and $\eta_2$, the detected fields
are given by
\begin{subequations}
\begin{align}
   d_1&= \sqrt{\eta_1} a_\mathrm{o} + i \sqrt{1-\eta_1} b_{\mathrm{LO1}},\label{eq:46} \\
   d_2&= \sqrt{\eta_2} c_\mathrm{o} + i \sqrt{1-\eta_2} b_{\mathrm{LO2}}, \label{eq:47}\\
   e_1&= \sqrt{\eta_1} b_{\mathrm{LO1}} + i \sqrt{1-\eta_1} a_\mathrm{o},\label{eq:48} \\
   e_2&= \sqrt{\eta_2} b_{\mathrm{LO2}} + i \sqrt{1-\eta_2} c_\mathrm{o},  \label{eq:49}
\end{align}
\end{subequations}
where $b_{1,2}$ are the fields of the local oscillators. With the definitions
given by Eq.~(\ref{eq:46}~-~\ref{eq:49}) and assuming that the LO state is
described by a coherent state $\bra b_{LO1}\ket=\beta_1 \exp \left[i \theta \right]$, we
can calculate
\begin{equation}
\label{eq:50}
\begin{split}
    \langle d_1^\dagger d_1\rangle& =  
    \left(1-\eta_1\right)  \langle b_{\mathrm{LO1}}^\dagger b_{\mathrm{LO1}}\rangle + \eta_1     
     \langle a^\dagger_\mathrm{o} a_\mathrm{o} \rangle -
      i \sqrt{\eta_1 \left(1-\eta_1\right)}  \left[ \langle b_{\mathrm{LO1}}^\dagger a_\mathrm{o} \rangle - \langle a ^\dagger_\mathrm{o} b_{\mathrm{LO1}}  \rangle \right] \\   
	&=\left(1-\eta_1\right)  \left| \beta_1 \right|^2 + \eta_1      \langle a^\dagger_\mathrm{o} a_\mathrm{o} \rangle +
       \sqrt{\eta_1 \left(1-\eta_1\right)} \left| \beta_1 \right| \langle X_a^\theta \rangle ,
\end{split}
\end{equation}
where
\begin{align*}
X_{\rm a}^\theta =
X^{\rm a}\left(\theta+\pi/2\right)= - i \left(a_\mathrm{o} \exp\left[-i \theta \right]
 - a^\dagger_\mathrm{o} \exp\left[i \theta \right] \right).
\end{align*}
Similarly one obtains
\begin{align}
\label{eq:51}
	\langle e_1^\dagger e_1\rangle& =  
    \eta_1  \left| \beta_1 \right|^2 + \left(1-\eta_1\right)   \langle a^\dagger_\mathrm{o} a_\mathrm{o} \rangle - \sqrt{\eta_1 \left(1-\eta_1\right)} \left| \beta_1 \right| \langle X_{\rm a}^\theta \rangle,
\end{align}
and analogously for detector $2$.

In addition to the intensities at the detectors D1, D2, E1, E2 we have to
evaluate the correlations among them. To this end we evaluate he full expression
for $\langle d_1^\dagger d_2^\dagger d_2 d_1\rangle$ which is given by
% Typo corrected: changed b_{\mathrm{LO1}} dagger to a dagger in the last term of the second to last row.
\begin{equation}
\label{eq:52}
\begin{split}
  R_{+\,+}\left(\theta,\phi\right)=& \langle d_1^\dagger d_2^\dagger d_2 d_1\rangle \\  
           =&\left(1-\eta_1\right) \left(1-\eta_2\right) \langle b_{\mathrm{LO1}}^\dagger b_{\mathrm{LO2}}^\dagger b_{\mathrm{LO2}} b_{\mathrm{LO1}} \rangle\\
         &+ i \sqrt{\eta_1\left(1-\eta_1\right)}(1-\eta_2) \left( \langle a^\dagger_\mathrm{o} b_{\mathrm{LO2}}^\dagger b_{\mathrm{LO2}} b_{\mathrm{LO1}} \rangle -
  \langle b_{\mathrm{LO1}}^\dagger b_{\mathrm{LO2}}^\dagger b_{\mathrm{LO2}} a_\mathrm{o}\rangle \right) \\
           &+ i \sqrt{\eta_2\left(1-\eta_2\right)}(1-\eta_1) \left( \langle b_{\mathrm{LO1}}^\dagger c^\dagger_\mathrm{o} b_{\mathrm{LO2}} b_{\mathrm{LO1}} \rangle -
  \langle b_{\mathrm{LO1}}^\dagger b_{\mathrm{LO2}}^\dagger c_\mathrm{o} b_{\mathrm{LO1}} \rangle \right) \\
           &+ \eta_1 (1-\eta_2) \langle a^\dagger_\mathrm{o} b_{\mathrm{LO2}}^\dagger b_{\mathrm{LO2}} a_\mathrm{o}\rangle  + \eta_2 (1-\eta_1) \langle b_{\mathrm{LO1}}^\dagger c^\dagger_\mathrm{o} c_\mathrm{o} b_{\mathrm{LO1}} \rangle \\
           &- \sqrt{\eta_1 \eta_2} \sqrt{\left(1-\eta_1\right)\left(1-\eta_2\right)}\left(\langle b_{\mathrm{LO1}}^\dagger
             b_{\mathrm{LO2}}^\dagger c_\mathrm{o} a_\mathrm{o} \rangle + \langle a^\dagger_\mathrm{o} c^\dagger_\mathrm{o} b_{\mathrm{LO2}} b_{\mathrm{LO1}}\rangle - \langle b_{\mathrm{LO1}}^\dagger c^\dagger_\mathrm{o} b_{\mathrm{LO2}} a_\mathrm{o}
             \rangle -\langle a^\dagger_\mathrm{o} b_{\mathrm{LO2}}^\dagger c_\mathrm{o} b_{\mathrm{LO1}} \rangle\right) \\
          &+  i \sqrt{\eta_1 \left(1-\eta_1 \right)} \eta_2 
                \left(\langle a^\dagger_\mathrm{o} c^\dagger_\mathrm{o} c_\mathrm{o} b_{\mathrm{LO1}} \rangle -\langle b_{\mathrm{LO1}}^\dagger c^\dagger_\mathrm{o} c_\mathrm{o} a_\mathrm{o}\rangle \right) \\
           &+ i \sqrt{\eta_2 \left(1-\eta_2 \right)} \eta_1 
                \left(\langle a^\dagger_\mathrm{o} c^\dagger_\mathrm{o} b_{\mathrm{LO2}} a_\mathrm{o}\rangle -\langle a^\dagger_\mathrm{o} b_{\mathrm{LO2}}^\dagger c_\mathrm{o} a_\mathrm{o}\rangle \right) \\ 
          &+\eta_1 \eta_2 \langle a^\dagger_\mathrm{o} c^\dagger_\mathrm{o} c_\mathrm{o} a_\mathrm{o}\rangle
\end{split}
\end{equation}
and, since we assume the LO to be in a coherent state, we have that
$b_{\mathrm{LO1}} \to \left|\beta_1\right| \exp \left[ i \theta \right] $,
$b_{\mathrm{LO2}} \to \left|\beta_2 \right| \exp \left[ i \phi \right] $, we get
% Typo corrected: exchanged the coefficients of the normal-ordered terms in row 3.
\begin{align}
\label{eq:53}
R_{+\,+}\left(\theta,\phi\right)=&\langle d_1^\dagger d_2^\dagger d_2 d_1\rangle \nonumber\\  
     =& \left(1-\eta_1\right) \left(1-\eta_2\right)  \left| \beta_1 \beta_2\right|^2 \nonumber\\ 
      &+ (1-\eta_2) \sqrt{\eta_1\left(1-\eta_1\right)} \left| \beta_2\right|^2 \left|\beta_1\right| \langle X^\theta_{\rm a}\rangle
      + (1-\eta_1) \sqrt{\eta_2\left(1-\eta_2\right)}   \left| \beta_1\right|^2 \left|\beta_2\right| \langle X^\phi_{\rm c}\rangle \nonumber\\   
     & +\sqrt{\eta_1 \eta_2} \sqrt{\left(1-\eta_1\right)\left(1-\eta_2\right)} \left| \beta_1 \beta_2 \right| \langle:X^\theta_{\rm a} X^\phi_{\rm c}:\rangle \nonumber\\   
     & +\eta_1 (1-\eta_2) \left| \beta_2\right|^2 \langle a^\dagger_\mathrm{o} a_\mathrm{o}\rangle 
        +\eta_2 (1-\eta_1) \left| \beta_1\right|^2 \langle c^\dagger_\mathrm{o} c_\mathrm{o}\rangle \nonumber\\   
     & +\eta_2 \sqrt{\eta_1 \left(1-\eta_1\right)} \left| \beta_1 \right| \langle :X^\theta_{\rm a} c^\dagger_\mathrm{o} c_\mathrm{o} : \rangle
        +\eta_1 \sqrt{\eta_2 \left(1-\eta_2\right)} \left| \beta_2 \right| \langle :X^\phi_{\rm c} a^\dagger_\mathrm{o} a_\mathrm{o} : \rangle \nonumber\\   
     & +\eta_1 \eta_2 \langle a^\dagger_\mathrm{o} c^\dagger_\mathrm{o} c_\mathrm{o} a_\mathrm{o}\rangle,
\end{align}
where with $\Braket{::}$ we denote normal ordering, i.e.
\begin{equation}
\label{eq:54}
\Braket{:X^{\theta}_{\rm a}  X^\phi_{\rm c}:}=-\Braket{a^\dagger_\mathrm{o} c^\dagger_\mathrm{o} \exp \left[ i\left(\theta+\phi\right)  \right] +c_\mathrm{o}\,a_\mathrm{o}\,
\exp \left[ -i\left(\theta+\phi\right) \right] -  c^\dagger_\mathrm{o} a_\mathrm{o} \, \exp \left[-i\left(\theta-\phi\right)\right]- a^\dagger_\mathrm{o} c_\mathrm{o}\,  \exp \left[ i\left(\theta-\phi\right) \right]} .
\end{equation}
The other terms are obtained replacing (where appropriate)
$\sqrt{\eta_1} \to i\sqrt{1-\eta_i}$ and $\sqrt{1-\eta_i} \to -i \sqrt{\eta_1}$
in Eqs.~\eqref{eq:52} and~\eqref{eq:53}. Using the expression of
$R_{\pm\,\pm}\left(\theta,\phi\right)$ given by Eq.~\eqref{eq:53} and assuming
50:50 beam splitters, i.e. $\eta_1=\eta_2=1/2$, the correlation coefficient
$E \left( \theta, \phi \right)$ in Eq.~\eqref{eq:28} of the main text can be written as
\begin{align}
\label{eq:55}
  E\left(\theta,\phi\right)=\frac{\left|\beta_1 \beta_2 \right| \Braket{:X^{\theta}_{\rm a}
  X^\phi_{\rm c}:}}{\left|\beta_1\right|^2\left|\beta_2\right|^2 + \left| \beta_1 \right|^2\Braket{c^\dagger_\mathrm{o} c_\mathrm{o}}+\left| \beta_2
  \right|^2\Braket{a^\dagger_\mathrm{o} a_\mathrm{o}}+ \Braket{a^\dagger_\mathrm{o} c^\dagger_\mathrm{o} c_\mathrm{o} a_\mathrm{o}}}.
\end{align}
In addition, it is possible to show \cite{Tan:1990ea} that the optimal value of the local oscillators for the violation of the Bell inequality is given by $\beta_1=\beta_2=\langle a^\dagger_\mathrm{o} c^\dagger_\mathrm{o} c_\mathrm{o} a_\mathrm{o}  \rangle^{1/4}$.
At this point, with the expression of the correlators given in Eqs.~(\ref{eq:50}-\ref{eq:54}),
we are in the position to express the correlation function $E\left(\theta,\phi\right)$ as
\begin{align}
\label{eq:56}
  E\left(\theta,\phi\right)= C \cos(\bar{\theta}-\bar{\phi}) + D \cos(\bar{\theta}+\bar{\phi}),
\end{align}
where
$\bar{\theta}-\bar{\phi}=\theta-\phi-\arg\langle a^\dagger_\mathrm{o}c_\mathrm{o}\rangle$,
$\bar{\theta}+\bar{\phi}=\theta+\phi-\arg\langle a^\dagger_\mathrm{o}c^\dagger_\mathrm{o}\rangle$.

The maxima of $S$ occur when $\bar{\theta}$ =0, $\bar{\phi}=-\zeta$,
$\bar{\theta}' =-\pi/2$ and $\bar{\phi}'=\zeta$ and with a maximum value is
given by
\begin{align}
S=2\sqrt{2} \sqrt{C^2+D^2} \sin(\zeta-\zeta_0),
\end{align}
where $\tan(\zeta_{0})=(C+D)/(C-D)$. The CHSH inequality, as
expressed in Eq.~\eqref{eq:27}, can be written as
\begin{align}
\mathcal{F} =C^2 + D^2<\dfrac{1}{2}
\end{align}
given  in Eq.~\eqref{eq:30} of the main text.

\section{Output field correlators}
\label{OFC}
In order to verify the violation of the CHSH inequality in the setup
described in the text, we evaluate 
\begin{align}
C=& \dfrac{2\left|\Braket{a^\dagger_\mathrm{o} c_\mathrm{o} }\right|}{2\sqrt{\Braket{a^\dagger_\mathrm{o}c^\dagger_\mathrm{o}  c_{\rm o} a_\mathrm{o}}}+\Braket{c^\dagger_o c_o} +\Braket{a^\dagger_\mathrm{o} a_\mathrm{o}}},\\
D=&\dfrac{2|\Braket{a_\mathrm{o} c_\mathrm{o} }|}{2\sqrt{\Braket{a^\dagger_\mathrm{o}c^\dagger_\mathrm{o}  c_{\rm o} a_\mathrm{o}}}+\Braket{c^\dagger_o c_o} +\Braket{a^\dagger_\mathrm{o} a_\mathrm{o}}}.
\end{align}
in the presence of two weak coherent drives for each cavity. In addition we
consider the possibility of the presence of thermal noise for the mechanics and
both cavities. The latter can be divided in "external" i.e. incoming through the
driving ports, or internal. In this case, we can write the input fields as
$a_{\rm i} =\alpha_{\rm i} + a_{\rm E}$ and
$c_{\rm i} =\chi_{\rm i} + c_{\rm E}$, where $\chi{\rm i}$ and $\alpha_{\rm i}$
represent the weak coherent drives, while $a_{\rm E}$ and $c_{\rm E}$ are the
operators associated to the ''external" thermal noise.

In this framework,  the correlations required to evaluate the CHSH inequality are given by   
\begin{align}
\Braket{a^\dagger_\mathrm{o} a_\mathrm{o}} =& \left| \Ad \right|^2 \left(\left|\ain\right|^2 + \nae\right) + \left|\Ax\right|^2 \left(\left|\cin\right|^2 + \nce + 1\right) + \Ad^* \Ax \ain^* \cin^* + \Ax^* \Ad \ain \cin \\
	&+ \left|\AdI\right|^2 \nai + \left|\AxI\right|^2 \left(\nci + 1\right) + \left|\Am\right|^2 \left(\nm + 1\right)\, , \nonumber \\
	\Braket{c^\dagger_\mathrm{o} c_\mathrm{o}} =& \left| \Cd \right|^2 \left(\left|\cin\right|^2 + \nce\right) + \left|\Cx\right|^2 \left(\left|\ain\right|^2 + \nae + 1\right) + \Cd^* \Cx \ain^* \cin^* + \Cx^* \Cd \ain \cin \\
	&+ \left|\CdI\right|^2 \nci + \left|\CxI\right|^2 \left(\nai + 1\right) + \left|\Cm\right|^2 \nm\,, \nonumber \\
	\Braket{a^\dagger_\mathrm{o} c_\mathrm{o}} =& \Ad^* \Cx \ain^{*2} + \left( \Ad^* \Cd + \Ax^* \Cx \right) \ain^* \cin + \Ax^* \Cd \cin^2, \\
	\Braket{a_\mathrm{o} c_\mathrm{o}} =& \Ad \Cx \left(\left|\ain\right|^2 + \nae + 1\right) + \Ax \Cd \left(\left|\cin\right|^2 + \nce\right) + \Ad \Cd \ain \cin + \Ax \Cx \ain^* \cin^* \\
	&+ \AdI \CxI \left(\nai + 1\right) + \AxI \CdI \nci + \Am \Cm \nm. \nonumber
\end{align}
Additionally the fourth order correlator is
\begin{align}
	\Braket{a^\dagger_\mathrm{o} c^\dagger_\mathrm{o} a_\mathrm{o} c_\mathrm{o}} =&\left|\Ad \Cx\right|^2 \left(\left|\ain\right|^4 + \left|\ain\right|^2 + 4\left|\ain\right|^2 \nae + n^2_\mathrm{e,a}\right) \\
	+& \left|\Ax \Cd\right|^2 \left(\left|\cin\right|^4 + 3\left|\cin\right|^2 + 4\left|\cin\right|^2 \nce + n^2_\mathrm{e,c} + 2\nce + 1\right) \nonumber\\
	+& \left|\Ad\right|^2 \left(\left|\ain\right| + \nae\right) \left[\left|\Cd\right|^2 \left(\left|\cin\right|^2 + \nce\right) + \left|\CdI\right|^2 \nci + \left|\CxI\right|^2 \left(\nai + 1\right) + \left|\Cm\right|^2 \nm\right] \nonumber \\
	+& \left|\Ax\right|^2 \left(\left|\cin\right| + \nce + 1\right) \left[\left|\Cx\right|^2 \left(\left|\ain\right|^2 + \nae + 1\right) + \left|\CdI\right|^2 \nci + \left|\CxI\right|^2 \left(\nai + 1\right) + \left|\Cm\right|^2 \nm\right] \nonumber \\
	+& \left|\AdI\right|^2 \nai \left[\left|\Cd\right|^2 \left(\left|\cin\right|^2 + \nce\right) + \left|\Cx\right|^2 \left(\left|\ain\right|^2 + \nae + 1\right) + \left|\CdI\right|^2 \nci + \left|\Cm\right|^2 \nm\right] \nonumber \\
	+& \left|\AxI\right|^2 \left(\nci + 1\right) \left[\left|\Cd\right|^2 \left(\left|\cin\right|^2 + \nce\right) + \left|\Cx\right|^2 \left(\left|\ain\right|^2 + \nae + 1\right) + \left|\CxI\right|^2 \left(\nai + 1\right) + \left|\Cm\right|^2 \nm\right] \nonumber \\
	+& \left|\Am\right|^2 \left(\nm + 1\right) \left[\left|\Cd\right|^2 \left(\left|\cin\right|^2 + \nce\right) + \left|\Cx\right|^2 \left(\left|\ain\right|^2 + \nae + 1\right) + \left|\CdI\right|^2 \nci + \left|\CxI\right|^2 \left(\nai + 1\right)\right] \nonumber \\
	+& \left|\AdI \CxI\right|^2 n^{\mathrm{I}2}_\mathrm{a} + \left|\AxI \CdI\right|^2 \left(n^{\mathrm{I}2}_\mathrm{c} + 1\right) + \left|\Am \Cm\right|^2 \left(\nm^2 + 2\nm + 1\right) \nonumber \\
	+& \Ad^* \Cd^* \ain^* \cin^* \left[\Ax \Cx \ain^* \cin^* + \AdI \CxI \nai + \AxI \CdI \left(\nci + 1\right) + \Am \Cm \left(\nm + 1\right)\right] \nonumber \\
	+& \Ad^* \Cx^* \left(\left|\ain\right|^2 + \nae\right) \left[\Ax \Cd \left(\left|\cin\right|^2 + \nce + 1\right) + \AdI \CxI \nai + \AxI \CdI \left(\nci + 1\right) + \Am \Cm \left(\nm + 1\right)\right] \nonumber \\
	+& \Ax^* \Cd^* \left(\left|\cin\right|^2 + \nce + 1\right) \left[\Ad \Cx \left(\left|\ain\right|^2 + \nae\right) + \AdI \CxI \nai + \AxI \CdI \left(\nci + 1\right) + \Am \Cm \left(\nm + 1\right)\right] \nonumber \\
	+& \Ax^* \Cx^* \ain \cin \left[\Ad \Cd \ain \cin + \AdI \CxI \nai + \AxI \CdI \left(\nci + 1\right) + \Am \Cm \left(\nm + 1\right)\right] \nonumber \\
	+& \AdI^* \CxI^* \nai \Big[\Ad \Cd \ain \cin + \Ad \Cx \left(\left|\ain\right|^2 + \nae\right) + \Ax \Cd \left(\left|\cin\right|^2 + \nce + 1\right) + \Ax \Cx \ain^* \cin^* + \nonumber \\
	&\qquad \qquad \qquad + \AxI \CdI \left(\nci + 1\right) + \Am \Cm \left(\nm + 1\right) \Big] \nonumber \\
	+& \AxI^* \CdI^* \left(\nci + 1\right) \Big[\Ad \Cd \ain \cin + \Ad \Cx \left(\left|\ain\right|^2 + \nae\right) + \Ax \Cd \left(\left|\cin\right|^2 + \nce + 1\right) + \Ax \Cx \ain^* \cin^* + \nonumber \\
	&\qquad \qquad \qquad + \AdI \CxI \nai + \Am \Cm \left(\nm + 1\right) \Big] \nonumber \\
	+& \Am^* \Cm^* \left(\nm + 1\right) \Big[\Ad \Cd \ain \cin + \Ad \Cx \left(\left|\ain\right|^2 + \nae\right) + \Ax \Cd \left(\left|\cin\right|^2 + \nce + 1\right) + \Ax \Cx \ain^* \cin^* + \nonumber \\
	&\qquad \qquad \qquad + \AdI \CxI \nai + \AxI \CdI \left(\nci + 1\right)\Big] \nonumber \\
	+& \left|\Ad\right|^2 \Cd^* \Cx \cin^* \left(\ain^* \left|\ain\right|^2 + 2 \ain^* \nae\right) + \left|\Ad\right|^2 \Cd \Cx^* \cin \left(\ain \left|\ain\right|^2 + 2 \ain \nae\right) \nonumber \\
	+& \left|\Ax\right|^2 \Cd^* \Cx \ain^* \left(\cin^* \left|\cin\right|^2 + 2 \cin^* \nce + 2 \cin^*\right) + \left|\Ax\right|^2 \Cd \Cx^* \ain \left(\cin \left|\cin\right|^2 + 2 \cin \nce + 2 \cin\right) \nonumber \\
	+& \Ad^* \Ax \left|\Cd\right|^2 \ain^* \left(\cin^* \left|\cin\right|^2 + 2 \cin^* \nce + \cin^*\right) + \Ad \Ax^* \left|\Cd\right|^2 \ain \left(\cin \left|\cin\right|^2 + 2 \cin \nce + \cin\right) \nonumber \\
	+& \Ad^* \Ax \left|\Cx\right|^2 \cin^* \left(\ain^* \left|\ain\right|^2 + 2 \ain^* \nae + \ain^*\right) + \Ad \Ax^* \left|\Cx\right|^2 \cin \left(\ain \left|\ain\right|^2 + 2 \ain \nae + \ain\right). \nonumber 
\end{align}

%\bibliographystyle{apsrev_abb}
%\bibliography{Bell}

\begin{thebibliography}{45}
\expandafter\ifx\csname natexlab\endcsname\relax\def\natexlab#1{#1}\fi
\expandafter\ifx\csname bibnamefont\endcsname\relax
  \def\bibnamefont#1{#1}\fi
\expandafter\ifx\csname bibfnamefont\endcsname\relax
  \def\bibfnamefont#1{#1}\fi
\expandafter\ifx\csname citenamefont\endcsname\relax
  \def\citenamefont#1{#1}\fi
\expandafter\ifx\csname url\endcsname\relax
  \def\url#1{\texttt{#1}}\fi
\expandafter\ifx\csname urlprefix\endcsname\relax\def\urlprefix{URL }\fi
\providecommand{\bibinfo}[2]{#2}
\providecommand{\eprint}[2][]{\url{#2}}

\bibitem[{\citenamefont{Einstein et~al.}(1935)\citenamefont{Einstein, Podolsky,
  and Rosen}}]{Einstein:1935hx}
\bibinfo{author}{\bibfnamefont{A.}~\bibnamefont{Einstein}},
  \bibinfo{author}{\bibfnamefont{B.}~\bibnamefont{Podolsky}}, \bibnamefont{and}
  \bibinfo{author}{\bibfnamefont{N.}~\bibnamefont{Rosen}},
  \bibinfo{journal}{Phys. Rev.} \textbf{\bibinfo{volume}{47}},
  \bibinfo{pages}{777} (\bibinfo{year}{1935}).

\bibitem[{\citenamefont{Bell}(1964)}]{Bell:1964wu}
\bibinfo{author}{\bibfnamefont{J.~S.} \bibnamefont{Bell}},
  \bibinfo{journal}{Physics} \textbf{\bibinfo{volume}{1}}, \bibinfo{pages}{195}
  (\bibinfo{year}{1964}).

\bibitem[{\citenamefont{Brunner et~al.}(2014)\citenamefont{Brunner, Cavalcanti,
  Pironio, Scarani, and Wehner}}]{Brunner:2014kr}
\bibinfo{author}{\bibfnamefont{N.}~\bibnamefont{Brunner}},
  \bibnamefont{et~al.}, \bibinfo{journal}{Rev. Mod. Phys.}
  \textbf{\bibinfo{volume}{86}}, \bibinfo{pages}{419} (\bibinfo{year}{2014}).

\bibitem[{\citenamefont{Ac{\'\i}n et~al.}(2007)\citenamefont{Ac{\'\i}n,
  Brunner, Gisin, Massar, Pironio, and Scarani}}]{Acin:2007db}
\bibinfo{author}{\bibfnamefont{A.}~\bibnamefont{Ac{\'\i}n}},
  \bibnamefont{et~al.}, \bibinfo{journal}{Phys. Rev. Lett.}
  \textbf{\bibinfo{volume}{98}}, \bibinfo{pages}{230501}
  (\bibinfo{year}{2007}).

\bibitem[{\citenamefont{Freedman and Clauser}(1972)}]{Freedman:1972ka}
\bibinfo{author}{\bibfnamefont{S.~J.} \bibnamefont{Freedman}} \bibnamefont{and}
  \bibinfo{author}{\bibfnamefont{J.~F.} \bibnamefont{Clauser}},
  \bibinfo{journal}{Phys. Rev. Lett.} \textbf{\bibinfo{volume}{28}},
  \bibinfo{pages}{938} (\bibinfo{year}{1972}).

\bibitem[{\citenamefont{Fry and Thompson}(1976)}]{Fry:1976hv}
\bibinfo{author}{\bibfnamefont{E.~S.} \bibnamefont{Fry}} \bibnamefont{and}
  \bibinfo{author}{\bibfnamefont{R.~C.} \bibnamefont{Thompson}},
  \bibinfo{journal}{Phys. Rev. Lett.} \textbf{\bibinfo{volume}{37}},
  \bibinfo{pages}{465} (\bibinfo{year}{1976}).

\bibitem[{\citenamefont{Aspect et~al.}(1982{\natexlab{a}})\citenamefont{Aspect,
  Dalibard, and Roger}}]{Aspect:1982ja}
\bibinfo{author}{\bibfnamefont{A.}~\bibnamefont{Aspect}},
  \bibinfo{author}{\bibfnamefont{J.}~\bibnamefont{Dalibard}}, \bibnamefont{and}
  \bibinfo{author}{\bibfnamefont{G.}~\bibnamefont{Roger}},
  \bibinfo{journal}{Phys. Rev. Lett.} \textbf{\bibinfo{volume}{49}},
  \bibinfo{pages}{1804} (\bibinfo{year}{1982}{\natexlab{a}}).

\bibitem[{\citenamefont{Aspect et~al.}(1982{\natexlab{b}})\citenamefont{Aspect,
  Grangier, and Roger}}]{Aspect:1982br}
\bibinfo{author}{\bibfnamefont{A.}~\bibnamefont{Aspect}},
  \bibinfo{author}{\bibfnamefont{P.}~\bibnamefont{Grangier}}, \bibnamefont{and}
  \bibinfo{author}{\bibfnamefont{G.}~\bibnamefont{Roger}},
  \bibinfo{journal}{Phys. Rev. Lett.} \textbf{\bibinfo{volume}{49}},
  \bibinfo{pages}{91} (\bibinfo{year}{1982}{\natexlab{b}}).

\bibitem[{\citenamefont{Weihs et~al.}(1998)\citenamefont{Weihs, Jennewein,
  Simon, Weinfurter, and Zeilinger}}]{Weihs:1998cc}
\bibinfo{author}{\bibfnamefont{G.}~\bibnamefont{Weihs}}, \bibnamefont{et~al.},
  \bibinfo{journal}{Phys. Rev. Lett.} \textbf{\bibinfo{volume}{81}},
  \bibinfo{pages}{5039} (\bibinfo{year}{1998}).

\bibitem[{\citenamefont{Rowe et~al.}(2001)\citenamefont{Rowe, Kielpinski,
  Meyer, Sackett, Itano, Monroe, and Wineland}}]{Rowe:2001ic}
\bibinfo{author}{\bibfnamefont{M.~A.} \bibnamefont{Rowe}},
  \bibnamefont{et~al.}, \bibinfo{journal}{Nature}
  \textbf{\bibinfo{volume}{409}}, \bibinfo{pages}{791} (\bibinfo{year}{2001}).

\bibitem[{\citenamefont{Matsukevich et~al.}(2008)\citenamefont{Matsukevich,
  Maunz, Moehring, Olmschenk, and Monroe}}]{Matsukevich:2008bm}
\bibinfo{author}{\bibfnamefont{D.~N.} \bibnamefont{Matsukevich}},
  \bibnamefont{et~al.}, \bibinfo{journal}{Phys. Rev. Lett.}
  \textbf{\bibinfo{volume}{100}}, \bibinfo{pages}{150404}
  (\bibinfo{year}{2008}).

\bibitem[{\citenamefont{Ansmann et~al.}(2009)\citenamefont{Ansmann, Wang,
  Bialczak, Hofheinz, Lucero, Neeley, O'Connell, Sank, Weides, Wenner
  et~al.}}]{Ansmann:2009eq}
\bibinfo{author}{\bibfnamefont{M.}~\bibnamefont{Ansmann}},
  \bibnamefont{et~al.}, \bibinfo{journal}{Nature}
  \textbf{\bibinfo{volume}{461}}, \bibinfo{pages}{504} (\bibinfo{year}{2009}).

\bibitem[{\citenamefont{Giustina et~al.}(2013)\citenamefont{Giustina, Mech,
  Ramelow, Wittmann, Kofler, Beyer, Lita, Calkins, Gerrits, Nam
  et~al.}}]{Giustina:2013jsa}
\bibinfo{author}{\bibfnamefont{M.}~\bibnamefont{Giustina}},
  \bibnamefont{et~al.}, \bibinfo{journal}{Nature}
  \textbf{\bibinfo{volume}{497}}, \bibinfo{pages}{227} (\bibinfo{year}{2013}).

\bibitem[{\citenamefont{Christensen et~al.}(2013)\citenamefont{Christensen,
  McCusker, Altepeter, Calkins, Gerrits, Lita, Miller, Shalm, Zhang, Nam
  et~al.}}]{Christensen:2013dn}
\bibinfo{author}{\bibfnamefont{B.~G.} \bibnamefont{Christensen}},
  \bibnamefont{et~al.}, \bibinfo{journal}{Phys. Rev. Lett.}
  \textbf{\bibinfo{volume}{111}}, \bibinfo{pages}{195} (\bibinfo{year}{2013}).

\bibitem[{\citenamefont{Hensen et~al.}(2015)\citenamefont{Hensen, Bernien,
  Dr{\'e}au, Reiserer, Kalb, Blok, Ruitenberg, Vermeulen, Schouten, Abell{\'a}n
  et~al.}}]{Hensen:2015dw}
\bibinfo{author}{\bibfnamefont{B.}~\bibnamefont{Hensen}}, \bibnamefont{et~al.},
  \bibinfo{journal}{Nature} \textbf{\bibinfo{volume}{526}},
  \bibinfo{pages}{682} (\bibinfo{year}{2015}).

\bibitem[{\citenamefont{Giustina et~al.}(2015)\citenamefont{Giustina,
  Versteegh, Wengerowsky, Handsteiner, Hochrainer, Phelan, Steinlechner,
  Kofler, Larsson, Abell{\'a}n et~al.}}]{Giustina:2015fc}
\bibinfo{author}{\bibfnamefont{M.}~\bibnamefont{Giustina}},
  \bibnamefont{et~al.}, \bibinfo{journal}{Phys. Rev. Lett.}
  \textbf{\bibinfo{volume}{115}}, \bibinfo{pages}{250401}
  (\bibinfo{year}{2015}).

\bibitem[{\citenamefont{Shalm et~al.}(2015)\citenamefont{Shalm, Meyer-Scott,
  Christensen, Bierhorst, Wayne, Stevens, Gerrits, Glancy, Hamel, Allman
  et~al.}}]{Shalm:2015cw}
\bibinfo{author}{\bibfnamefont{L.~K.} \bibnamefont{Shalm}},
  \bibnamefont{et~al.}, \bibinfo{journal}{Phys. Rev. Lett.}
  \textbf{\bibinfo{volume}{115}}, \bibinfo{pages}{250402}
  (\bibinfo{year}{2015}).

\bibitem[{\citenamefont{Thearle et~al.}(2018)\citenamefont{Thearle, Janousek,
  Armstrong, Hosseini, Sch{\"u}nemann~Mraz, Assad, Symul, James, Huntington,
  Ralph et~al.}}]{Thearle:2018ib}
\bibinfo{author}{\bibfnamefont{O.}~\bibnamefont{Thearle}},
  \bibnamefont{et~al.}, \bibinfo{journal}{Phys. Rev. Lett.}
  \textbf{\bibinfo{volume}{120}}, \bibinfo{pages}{040406}
  (\bibinfo{year}{2018}).

\bibitem[{\citenamefont{Huntington and Ralph}(2001)}]{Huntington:2001kr}
\bibinfo{author}{\bibfnamefont{E.~H.} \bibnamefont{Huntington}}
  \bibnamefont{and} \bibinfo{author}{\bibfnamefont{T.~C.} \bibnamefont{Ralph}},
  \bibinfo{journal}{Phys. Rev. A} \textbf{\bibinfo{volume}{65}},
  \bibinfo{pages}{012306} (\bibinfo{year}{2001}).

\bibitem[{\citenamefont{Ou and Mandel}(1988)}]{Ou:1988gk}
\bibinfo{author}{\bibfnamefont{Z.~Y.} \bibnamefont{Ou}} \bibnamefont{and}
  \bibinfo{author}{\bibfnamefont{L.}~\bibnamefont{Mandel}},
  \bibinfo{journal}{Phys. Rev. Lett.} \textbf{\bibinfo{volume}{61}},
  \bibinfo{pages}{50} (\bibinfo{year}{1988}).

\bibitem[{\citenamefont{Hammerer et~al.}(2014)\citenamefont{Hammerer, Genes,
  Vitali, Tombesi, Milburn, Simon, and Bouwmeester}}]{Hammerer:2014vu}
\bibinfo{author}{\bibfnamefont{K.}~\bibnamefont{Hammerer}},
  \bibnamefont{et~al.}, in \emph{\bibinfo{booktitle}{Cavity Optomechanics:
  Nano- and Micromechanical Resonators Interacting with Light}}
  (\bibinfo{publisher}{Springer Berlin Heidelberg}, \bibinfo{address}{Berlin,
  Heidelberg}, \bibinfo{year}{2014}), pp. \bibinfo{pages}{25--56}.

\bibitem[{\citenamefont{Aspelmeyer et~al.}(2014)\citenamefont{Aspelmeyer,
  Kippenberg, and Marquardt}}]{Aspelmeyer:2014ce}
\bibinfo{author}{\bibfnamefont{M.}~\bibnamefont{Aspelmeyer}},
  \bibinfo{author}{\bibfnamefont{T.~J.} \bibnamefont{Kippenberg}},
  \bibnamefont{and}
  \bibinfo{author}{\bibfnamefont{F.}~\bibnamefont{Marquardt}},
  \bibinfo{journal}{Rev. Mod. Phys.} \textbf{\bibinfo{volume}{86}},
  \bibinfo{pages}{1391} (\bibinfo{year}{2014}).

\bibitem[{\citenamefont{Abbott et~al.}(2016)\citenamefont{Abbott, Abbott,
  Abbott, Abernathy, Acernese, Ackley, Adams, Adams, Addesso, Adhikari
  et~al.}}]{Abbott:2016ki}
\bibinfo{author}{\bibfnamefont{B.~P.} \bibnamefont{Abbott}},
  \bibnamefont{et~al.}, \bibinfo{journal}{Phys. Rev. Lett.}
  \textbf{\bibinfo{volume}{116}}, \bibinfo{pages}{061102}
  (\bibinfo{year}{2016}).

\bibitem[{\citenamefont{Pikovski et~al.}(2012)\citenamefont{Pikovski, Vanner,
  Aspelmeyer, Kim, and Brukner}}]{Pikovski:2012hp}
\bibinfo{author}{\bibfnamefont{I.}~\bibnamefont{Pikovski}},
  \bibnamefont{et~al.}, \bibinfo{journal}{Nat. Phys.}
  \textbf{\bibinfo{volume}{8}}, \bibinfo{pages}{393} (\bibinfo{year}{2012}).

\bibitem[{\citenamefont{Wang and Clerk}(2013)}]{Wang:2013hk}
\bibinfo{author}{\bibfnamefont{Y.-D.} \bibnamefont{Wang}} \bibnamefont{and}
  \bibinfo{author}{\bibfnamefont{A.~A.} \bibnamefont{Clerk}},
  \bibinfo{journal}{Phys. Rev. Lett.} \textbf{\bibinfo{volume}{110}},
  \bibinfo{pages}{253601} (\bibinfo{year}{2013}).

\bibitem[{\citenamefont{Woolley and Clerk}(2014)}]{Woolley:2014he}
\bibinfo{author}{\bibfnamefont{M.~J.} \bibnamefont{Woolley}} \bibnamefont{and}
  \bibinfo{author}{\bibfnamefont{A.~A.} \bibnamefont{Clerk}},
  \bibinfo{journal}{Phys. Rev. A} \textbf{\bibinfo{volume}{89}},
  \bibinfo{pages}{063805} (\bibinfo{year}{2014}).

\bibitem[{\citenamefont{Massel}(2017)}]{Massel:2017jx}
\bibinfo{author}{\bibfnamefont{F.}~\bibnamefont{Massel}},
  \bibinfo{journal}{Phys. Rev. A} \textbf{\bibinfo{volume}{95}},
  \bibinfo{pages}{063816} (\bibinfo{year}{2017}).

\bibitem[{\citenamefont{Riedinger et~al.}(2017)\citenamefont{Riedinger,
  Wallucks, Marinkovic, L{\"o}schnauer, Aspelmeyer, Hong, and
  Gr{\"o}blacher}}]{Riedinger:2017wa}
\bibinfo{author}{\bibfnamefont{R.}~\bibnamefont{Riedinger}},
  \bibnamefont{et~al.} (\bibinfo{year}{2017}), \eprint{1710.11147}.

\bibitem[{\citenamefont{Ockeloen-Korppi
  et~al.}(2018)\citenamefont{Ockeloen-Korppi, Damsk{\"a}gg, Pirkkalainen,
  Asjad, Clerk, Massel, Woolley, and
  Sillanp{\"a}{\"a}}}]{OckeloenKorppi:2018ks}
\bibinfo{author}{\bibfnamefont{C.~F.} \bibnamefont{Ockeloen-Korppi}},
  \bibnamefont{et~al.}, \bibinfo{journal}{Nature}
  \textbf{\bibinfo{volume}{556}}, \bibinfo{pages}{478} (\bibinfo{year}{2018}).

\bibitem[{\citenamefont{Clauser et~al.}(1969)\citenamefont{Clauser, Horne,
  Shimony, and Holt}}]{Clauser:1969ff}
\bibinfo{author}{\bibfnamefont{J.~F.} \bibnamefont{Clauser}},
  \bibnamefont{et~al.}, \bibinfo{journal}{Phys. Rev. Lett.}
  \textbf{\bibinfo{volume}{23}}, \bibinfo{pages}{880} (\bibinfo{year}{1969}).

\bibitem[{\citenamefont{Barzanjeh et~al.}(2012)\citenamefont{Barzanjeh, Abdi,
  Milburn, Tombesi, and Vitali}}]{Barzanjeh:2012ez}
\bibinfo{author}{\bibfnamefont{S.}~\bibnamefont{Barzanjeh}},
  \bibnamefont{et~al.}, \bibinfo{journal}{Phys. Rev. Lett.}
  \textbf{\bibinfo{volume}{109}}, \bibinfo{pages}{130503}
  (\bibinfo{year}{2012}).

\bibitem[{\citenamefont{Paternostro et~al.}(2007)\citenamefont{Paternostro,
  Vitali, Gigan, Kim, Brukner, Eisert, and Aspelmeyer}}]{Paternostro:2007hm}
\bibinfo{author}{\bibfnamefont{M.}~\bibnamefont{Paternostro}},
  \bibnamefont{et~al.}, \bibinfo{journal}{Phys. Rev. Lett.}
  \textbf{\bibinfo{volume}{99}}, \bibinfo{pages}{250401}
  (\bibinfo{year}{2007}).

\bibitem[{\citenamefont{Ockeloen-Korppi
  et~al.}(2016)\citenamefont{Ockeloen-Korppi, Damsk{\"a}gg, Pirkkalainen,
  Heikkil{\"a}, Massel, and Sillanp{\"a}{\"a}}}]{OckeloenKorppi:2016ke}
\bibinfo{author}{\bibfnamefont{C.~F.} \bibnamefont{Ockeloen-Korppi}},
  \bibnamefont{et~al.}, \bibinfo{journal}{Phys. Rev. X}
  \textbf{\bibinfo{volume}{6}}, \bibinfo{pages}{041024} (\bibinfo{year}{2016}).

\bibitem[{\citenamefont{Vivoli et~al.}(2016)\citenamefont{Vivoli, Barnea,
  Galland, and Sangouard}}]{Vivoli:2016fp}
\bibinfo{author}{\bibfnamefont{V.~C.} \bibnamefont{Vivoli}},
  \bibnamefont{et~al.}, \bibinfo{journal}{Phys. Rev. Lett.}
  \textbf{\bibinfo{volume}{116}}, \bibinfo{pages}{070405}
  (\bibinfo{year}{2016}).

\bibitem[{\citenamefont{Hofer et~al.}(2016)\citenamefont{Hofer, Lehnert, and
  Hammerer}}]{Hofer:2016id}
\bibinfo{author}{\bibfnamefont{S.~G.} \bibnamefont{Hofer}},
  \bibinfo{author}{\bibfnamefont{K.~W.} \bibnamefont{Lehnert}},
  \bibnamefont{and} \bibinfo{author}{\bibfnamefont{K.}~\bibnamefont{Hammerer}},
  \bibinfo{journal}{Phys. Rev. Lett.} \textbf{\bibinfo{volume}{116}},
  \bibinfo{pages}{070406} (\bibinfo{year}{2016}).

\bibitem[{\citenamefont{Tan et~al.}(1990)\citenamefont{Tan, Holland, and
  Walls}}]{Tan:1990ea}
\bibinfo{author}{\bibfnamefont{S.~M.} \bibnamefont{Tan}},
  \bibinfo{author}{\bibfnamefont{M.~J.} \bibnamefont{Holland}},
  \bibnamefont{and} \bibinfo{author}{\bibfnamefont{D.~F.} \bibnamefont{Walls}},
  \bibinfo{journal}{Optics Communications} \textbf{\bibinfo{volume}{77}},
  \bibinfo{pages}{285} (\bibinfo{year}{1990}).

\bibitem[{\citenamefont{Tan et~al.}(1991)\citenamefont{Tan, Walls, and
  Collett}}]{Tan:1991cg}
\bibinfo{author}{\bibfnamefont{S.~M.} \bibnamefont{Tan}},
  \bibinfo{author}{\bibfnamefont{D.~F.} \bibnamefont{Walls}}, \bibnamefont{and}
  \bibinfo{author}{\bibfnamefont{M.~J.} \bibnamefont{Collett}},
  \bibinfo{journal}{Phys. Rev. Lett.} \textbf{\bibinfo{volume}{66}},
  \bibinfo{pages}{252} (\bibinfo{year}{1991}).

\bibitem[{\citenamefont{Law}(1995)}]{Law:1995it}
\bibinfo{author}{\bibfnamefont{C.~K.} \bibnamefont{Law}},
  \bibinfo{journal}{Phys. Rev. A} \textbf{\bibinfo{volume}{51}},
  \bibinfo{pages}{2537} (\bibinfo{year}{1995}).

\bibitem[{\citenamefont{Genes et~al.}(2009)\citenamefont{Genes, Mari, Vitali,
  and Tombesi}}]{Genes:2009cb}
\bibinfo{author}{\bibfnamefont{C.}~\bibnamefont{Genes}}, \bibnamefont{et~al.},
  \bibinfo{journal}{Advances in atomic, molecular, and optical physics}
  \textbf{\bibinfo{volume}{57}}, \bibinfo{pages}{33} (\bibinfo{year}{2009}).

\bibitem[{\citenamefont{Milburn and Woolley}(2011)}]{Milburn:2012cu}
\bibinfo{author}{\bibfnamefont{G.~J.} \bibnamefont{Milburn}} \bibnamefont{and}
  \bibinfo{author}{\bibfnamefont{M.~J.} \bibnamefont{Woolley}},
  \bibinfo{journal}{Acta Physica Slovaca} \textbf{\bibinfo{volume}{61}},
  \bibinfo{pages}{483} (\bibinfo{year}{2011}).

\bibitem[{\citenamefont{Bowen and Milburn}(2015)}]{Bowen:2015gt}
\bibinfo{author}{\bibfnamefont{W.~P.} \bibnamefont{Bowen}} \bibnamefont{and}
  \bibinfo{author}{\bibfnamefont{G.~J.} \bibnamefont{Milburn}},
  \emph{\bibinfo{title}{{Quantum Optomechanics}}} (\bibinfo{publisher}{CRC
  Press}, \bibinfo{year}{2015}).

\bibitem[{\citenamefont{Walls and Milburn}(2008)}]{Walls:2008em}
\bibinfo{author}{\bibfnamefont{D.~F.} \bibnamefont{Walls}} \bibnamefont{and}
  \bibinfo{author}{\bibfnamefont{G.~J.} \bibnamefont{Milburn}},
  \emph{\bibinfo{title}{{Quantum optics}}} (\bibinfo{publisher}{Springer Berlin
  Heidelberg}, \bibinfo{address}{Berlin, Heidelberg}, \bibinfo{year}{2008}).

\bibitem[{\citenamefont{Vidick and Wehner}(2011)}]{Vidick:2011bk}
\bibinfo{author}{\bibfnamefont{T.}~\bibnamefont{Vidick}} \bibnamefont{and}
  \bibinfo{author}{\bibfnamefont{S.}~\bibnamefont{Wehner}},
  \bibinfo{journal}{Phys. Rev. A} \textbf{\bibinfo{volume}{83}},
  \bibinfo{pages}{195} (\bibinfo{year}{2011}).

\bibitem[{\citenamefont{Junge and Palazuelos}(2011)}]{Junge:2011fv}
\bibinfo{author}{\bibfnamefont{M.}~\bibnamefont{Junge}} \bibnamefont{and}
  \bibinfo{author}{\bibfnamefont{C.}~\bibnamefont{Palazuelos}},
  \bibinfo{journal}{Communications in Mathematical Physics}
  \textbf{\bibinfo{volume}{306}}, \bibinfo{pages}{695} (\bibinfo{year}{2011}).

\bibitem[{\citenamefont{Vallone et~al.}(2014)\citenamefont{Vallone, Lima,
  G{\'o}mez, Ca{\~n}as, Larsson, Mataloni, and Cabello}}]{Vallone:2014id}
\bibinfo{author}{\bibfnamefont{G.}~\bibnamefont{Vallone}},
  \bibnamefont{et~al.}, \bibinfo{journal}{Phys. Rev. A}
  \textbf{\bibinfo{volume}{89}}, \bibinfo{pages}{195} (\bibinfo{year}{2014}).

\end{thebibliography}

\end{document}